\documentclass[12pt,leqno]{article}
\usepackage{amssymb}
\usepackage[mathscr]{eucal}
\usepackage{amsmath,amssymb,latexsym,theorem,bbm}
\usepackage{amsmath,amssymb,latexsym,theorem}
\usepackage{multirow}
\usepackage{color,url}
\usepackage{graphicx}
\usepackage{appendix}
\usepackage{subcaption}

\newcommand{\NN}{\mathbb{N}}
\newcommand{\QQ}{\mathbb{Q}}
\newcommand{\RR}{\mathbb{R}}

\newcommand{\bone}{{\boldsymbol{1}}}

\newcommand{\cD}{{\mathcal D}}

\newcommand{\cF}{{\mathcal F}}

\newcommand{\cN}{{\mathcal N}}

\newcommand{\dd}{\mathrm{d}}
\newcommand{\ee}{\mathrm{e}}

\newcommand{\VaR}{\operatorname{ \tt{VaR} } }
\newcommand{\gVaR}{\operatorname{ \tt{gVaR} } }
\newcommand{\ES}{\operatorname{ \tt{ES}} }
\newcommand{\PELVE}{\operatorname{ \tt{PELVE}} }
\newcommand{\PELVES}{\operatorname{ \tt{PELVE}_2} }
\newcommand{\TGINI}{\operatorname{ \tt{TGini}} }
\newcommand{\GS}{\operatorname{ \tt{GS}} }

\newcommand{\EE}{\operatorname{\mathbb{E}}}

\newcommand{\PP}{\operatorname{\mathbb{P}}}

\newcommand{\range}{\operatorname{Range}}

\newcommand{\vare}{\varepsilon}

\renewcommand{\mid}{\,|\,}

\renewcommand{\leq}{\leqslant}
\renewcommand{\geq}{\geqslant}

\newcommand{\distr}{\stackrel{\cD}{\longrightarrow}}

\newcommand{\bbone}{\mathbbm{1}}

\newcommand{\proofend}{\hfill\mbox{$\Box$}}

\numberwithin{equation}{section}

\theoremstyle{change} \theorembodyfont{\em}
\newtheorem{Lem}{Lemma.}[section]
\newtheorem{Thm}[Lem]{Theorem.}
\newtheorem{Pro}[Lem]{Proposition.}

\newtheorem{Def}[Lem]{Definition.}

\theorembodyfont{\rm}
\newtheorem{Rem}[Lem]{Remark.}
\newtheorem{Ex}[Lem]{Example.}

\begin{document}

\begin{center}
 {\bfseries\Large Probability equivalent level of Value at Risk\\[1mm] and higher-order Expected Shortfalls}

\vspace*{3mm}

{\sc\large
 M\'aty\'as $\text{Barczy}^{*,\diamond}$,
  Fanni K. $\text{Ned\'enyi}^{*}$,
  \ L\'aszl\'o $\text{S\"ut\H o}^{**}$ }

\end{center}

\vskip0.2cm

\noindent
 * ELKH-SZTE Analysis and Applications Research Group,
   Bolyai Institute, University of Szeged,
   Aradi v\'ertan\'uk tere 1, H--6720 Szeged, Hungary.

\noindent
  ** Former master student of Bolyai Institute, University of Szeged,
     Aradi v\'ertan\'uk tere 1, H-6720 Szeged, Hungary.

\noindent e-mails: barczy@math.u-szeged.hu (M. Barczy),
                   nfanni@math.u-szeged.hu (F. K. Ned\'enyi),
                   suto71528@gmail.com (L. S\"ut\H o).

\noindent $\diamond$ Corresponding author.

\vskip0.5cm

{\renewcommand{\thefootnote}{}
\footnote{M\'aty\'as Barczy was supported by the project TKP2021-NVA-09.
Project no.\ TKP2021-NVA-09 has been implemented with the support
 provided by the Ministry of Innovation and Technology of Hungary from the National Research, Development and Innovation Fund,
 financed under the TKP2021-NVA funding scheme.
Fanni K. Ned\'enyi was supported by the \'UNKP-22-4 New National Excellence Program of the Ministry for Culture
 and Innovation from the source of the National Research, Development and Innovation Fund. }
 }

\vspace*{-15mm}

\begin{abstract}
We investigate the probability equivalent level of Value at Risk and \ $n^{\mathrm{th}}$-order Expected Shortfall (called \ $\PELVE_n$),
 which can be considered as a variant of the notion of the probability equivalent level of Value at Risk and Expected Shortfall
 (called \ $\PELVE$) due to Li and Wang (2022).
We study the finiteness, uniqueness and several properties of \ $\PELVE_n$, \ we calculate \ $\PELVE_n$ \ of some notable distributions,
 \ $\PELVES$ \ of a random variable having generalized Pareto excess distribution, and we describe
 the asymptotic behaviour of \ $\PELVES$ of regularly varying distributions as the level tends to \ $0$.
Some properties of \ $n^{\mathrm{th}}$-order Expected Shortfall are also investigated.
Among others, it turns out that the Gini Shortfall at some level \ $p\in[0,1)$ \ corresponding to a (loading) parameter $\lambda\geq 0$
 \ is the linear combination of the Expected Shortfall at level \ $p$ \
 and the \ $2^{\mathrm{nd}}$-order Expected Shortfall at level \ $p$ \ with coefficients \ $1-2\lambda$ \
 and \ $2\lambda$, \ respectively.
\end{abstract}

\section{Introduction}
\label{section_intro}

The Fundamental Review of the Trading Book (FRTB) was introduced by the Basel Committee on Banking Supervision
in the years following the Global Financial Crisis of 2007-2009.
FRTB is expected to make a complete revision of the approach to calculating
risk-based capital requirements for investments.
It was originally supposed to be implemented in January 2023,
but a new starting date, January 2025, has been recently announced.
Value at Risk \ ($\VaR$, \ see Definition \ref{Def_VaR}) and Expected Shortfall \ ($\ES$, \ see Definition \ref{Def_ES_n}) are popular risk measures
 used to measure portfolio risk.
According to FRTB, the banks are supposed to use \ $\ES$ \ at the level \ $0.975$ \ instead of $\VaR$ at the level \ $0.99$ \
 for the bank-wide internal models to determine market risk capital requirements.

Motivated by the FRTB, Li and Wang \cite{LiWan} have recently introduced the notion of probability equivalent level of \ $\VaR$ \
 and \ $\ES$ \ ($\PELVE$, \ see Definition \ref{Def_PELVE}).
Roughly speaking, for an integrable random variable \ $X$ \ and \ $\vare\in(0,1)$, \ the \ $\PELVE$ of \ $X$ \ at the level
 \ $\vare$ \ is the infimum of those values \ $c\in[1,\frac{1}{\vare}]$ \ for which the \ $\ES$ \ of \ $X$ \ at level \ $1-c\vare$ \
 is less than or equal to the \ $\VaR$ \ of \ $X$ \ at the level \ $1-\vare$.
One can see that the level \ $\vare=0.01$ \ corresponds to the replacement of \ $\VaR$ \ at the level \ $0.99$ \ with the \ $\ES$ \
 at some appropriate level, which has particular importance due to the FRTB.

Very recently, Fiori and Rosazza Gianin \cite{FioGia} have proposed {two generalizations} of \ $\PELVE$.
As a first generalization, they have replaced the pair \ $(\VaR,\ES)$ \
 in the definition of \ $\PELVE$ \ with a general pair of monotone risk measures \ $(\varrho,\widetilde\varrho)$, \
 where \ $\widetilde\varrho$ \ is obtained from \ $\varrho$ \ by integration similarly as \ $\ES$ \ can be obtained from
 \ $\VaR$ \ by integration, for more details, see Definition \ref{Def_Fiori_Gianin}.
As a special case of their generalization, Fiori and Rosazza Gianin \cite{FioGia} have also considered the so-called
 conditional \ $\PELVE$, \ where \ $\varrho$ \ is chosen to be the \ $\ES$.
As a second generalization, Fiori and Rosazza Gianin \cite[Definition 2]{FioGia}
 have introduced the so-called distorted $\PELVE$ associated to a family of distortions, for more details, see Remark \ref{Rem3}.

Both the \ $\PELVE$ \ due to Li and Wang \cite{LiWan} and its generalizations due to Fiori and Rosazza Gianin \cite{FioGia}
 are defined under the minimal assumption that the random variable representing the risk has a finite first moment.
These risk measures enjoy satisfactory invariance and ordering properties,
 and they have interesting links to the tail index of a regularly varying random variable.
Fiori and Rosazza Gianin \cite[Section 4]{FioGia} have also studied \ $\PELVE$ \ and conditional \ $\PELVE$ \ of
 a random variable having generalized Pareto excess distribution.

In the present paper we study the probability equivalent level of \ $\VaR$ \ and
 a higher-order \ $\ES$ \ (see Definition \ref{Def_ES_n}), i.e.,
 we replace \ $\ES$ \ with a higher-order \ $\ES$ \ in the definition of \ $\PELVE$ \ due to Li and Wang \cite{LiWan}.
We note that this variant of PELVE is a special case of the newly introduced notion, called distorted PELVE,
 in Fiori and Rosazza Gianin \cite[Definition 2]{FioGia} (for more details, see Remark \ref{Rem3}).
In Appendix \ref{App_2ndES_GiniES}, we point out that the Gini Shortfall at some level \ $p\in[0,1)$ \ corresponding to a (loading) parameter $\lambda\geq 0$,
 introduced in Furman et al.\ \cite[formula (4.1)]{FurWanZit},
 is the linear combination of the Expected Shortfall at level \ $p$ \
 and the \ $2^{\mathrm{nd}}$-order Expected Shortfall at level \ $p$ \ with coefficients \ $1-2\lambda$ \
 and \ $2\lambda$, \ respectively.
This underlines the importance of studying the previously mentioned variant of \ $\PELVE$ \ in a more detailed way.

Let \ $\NN$ \ and \ $\RR$ \ denote the set of positive integers and real numbers, respectively.
For a function \ $f:\RR\to\RR$, \ its range \ $\{f(x): x\in\RR\}$ \ is denoted by \ $\range(f)$.
\ The random variables will be defined on an appropriate probability space \ $(\Omega,\cF,\PP)$.
\ The distribution function of a random variable \ $X:\Omega\to\RR$ \ is given by \ $F_X:\RR\to[0,1]$, \ $F_X(x):=\PP(X\leq x)$, \ $x\in\RR$.
\ The set of random variables \ $X$ \ satisfying \ $\EE(\vert X\vert)<\infty$ \ is denoted by \ $L^1$.
\ Convergence in distribution is denoted by \ $\distr$.

First, we recall the notion of Value at Risk.

\begin{Def}\label{Def_VaR}
Let \ $X$ \ be a random variable.
The Value at Risk of \ $X$ \ at a level \ $p \in [0,1]$ \ is defined by
 \[
  \VaR_X(p):=\inf\{x\in\RR : F_X(x)\geq p\},
 \]
 with the convention \ $\inf\emptyset:=\infty$.
\end{Def}

Note that \ $\VaR_X(p)$ \ is also called the (lower) quantile or a generalized inverse of \ $X$ \ at a level \ $p \in (0,1)$.
\ One may think about \ $X$ \ as the loss and profit of some financial position at a given time point,
 and, using actuarial notation, positive values of \ $X$ \ represent losses, while negative values profit.
For each \ $p\in(0,1)$, \ $\VaR_X(p)$ \ is the smallest value \ $x$ \ such that the probability
 of a loss \ $X$ \ greater than \ $x$ \ is at most \ $1-p$, \ or, roughly speaking,
 \ $\VaR_X(p)$ \ is the loss that is likely to be exceeded only \ $(1-p)100\%$ of the time.
While \ $\VaR$ \ is widely used and easy to compute, it has no information on the magnitude of the biggest \ $(1-p)100\%$
 of the losses.
Note also that \ $\VaR_X(0)=-\infty$ \ for any random variable \ $X$.

Next, we recall the notion of \ $n^{\mathrm{th}}$-order Expected Shortfall due to Fuchs et al.\ \cite[Example 2, part (4)]{FucSchSch}.

\begin{Def}\label{Def_ES_n}
Let \ $X$ \ be a random variable such that \ $X\in L^1$, \ and let \ $n\in\NN$.
\ The \ $n^{\mathrm{th}}$-order Expected Shortfall of \ $X$ \ at a level \ $p\in[0,1)$ \ is defined by
  \[
   \ES_{X,n}(p):=\frac{n}{1-p}\int_p^1 \left( \frac{s-p}{1-p} \right)^{n-1} \VaR_X(s) \,\dd s.
  \]
\end{Def}

Note that $\ES_{X,n}(p)$ \ can be written in the form
 \ $\ES_{X,n}(p)= \int_p^1 \frac{n(s-p)^{n-1}}{(1-p)^{n-1}} \VaR_X(s) \,\dd s$,
 \ where the function \ $\frac{n(s-p)^{n-1}}{(1-p)^{n-1}}$, $s\in[p,1]$,
 can be considered as a weight function with integral \ $1$ \ on $[p,1]$ \
 such that higher losses are weighted higher.

In the next remark we recall some basic properties of higher-order Expected Shortfalls.
For some further properties of higher-order Expected Shortfalls, see Appendix \ref{App_2ndES_prop}.

\begin{Rem}\label{Rem4}
(i). For \ $X\in L^1$, \ $n\in\NN$, \ and \ $p\in[0,1)$, \ we have \ $\ES_{X,n}(p)\in\RR$, \
 see Lemma \ref{Lem_ESn_vegesseg}.
Note also that the first order Expected Shortfall coincides with the usual Expected Shortfall (also called Conditional Value at Risk),
 so \ $\ES_{X,1}$ \ is simply denoted by \ $\ES_X$.

(ii). By Lemma 2 in Fuchs et al.\ \cite{FucSchSch}, the \ $n^{\mathrm{th}}$-order Expected Shortfall
  is monotone (in the sense that if $X\leq Y$, \ $X,Y\in L^1$, \ then \ $\ES_{X,n}(p)\leq \ES_{Y,n}(p)$, \ $p\in[0,1)$), positive homogeneous
  and translation invariant.
  Further, using that the distortion function corresponding to the \ $n^{\mathrm{th}}$-order Expected Shortfall (see the function \ $h_p$ \
  in the proof of Lemma \ref{Lem_ESn_vegesseg})
  is convex, we get that the \ $n^{\mathrm{th}}$-order Expected Shortfall  is subadditive, see Fuchs et al.\ \cite[Example 1/(4) and Theorem 4]{FucSchSch}.
 The above mentioned properties of the \ $n^{\mathrm{th}}$-order Expected Shortfall also follow from Proposition 2 and Theorem 3
  in Wang et al.\ \cite{WanWanWei}.
  All in all, the \ $n^{\mathrm{th}}$-order Expected Shortfall is a coherent risk measure on \ $L^1$.

(iii). For \ $X\in L^1$ \ and \ $n_1,n_2\in\NN$ \ with \ $n_1\leq n_2$, \ we have
 \begin{align}\label{help_ES2_ES}
   \VaR_X(p)\leq \ES_X(p), \qquad p\in [0,1), \qquad \text{and} \qquad \ES_{X,n_1}(p) \leq \ES_{X,n_2}(p),\qquad p\in[0,1).
 \end{align}
 where the second inequality is a consequence of Fuchs et al.\ \cite[Corollary 4, part (1)]{FucSchSch}.
Indeed,
 \[
   \ES_{X,n}(p) = \int_0^1 \VaR_X(s)\,\dd D_{n,p}(s),\qquad n\in\NN,\;\; p\in[0,1),
 \]
 where $D_{n,p}(s):=\big(\frac{s-p}{1-p}\big)^n \bone_{[p,1]}(s)$, \ $s\in[0,1]$, and we have that $D_{n_2,p}(s)\leq D_{n_1,p}(s)$, $s\in[0,1]$
 for each $n_1,n_2\in\NN$ with $n_1\leq n_2$.
In particular, we have that $\ES_X(p) \leq \ES_{X,n}(p)$, $n\in\NN$, $p\in[0,1)$.
The second inequality in \eqref{help_ES2_ES} also follow{} by Wang et al.\ \cite[part (i) of Proposition 2]{WanWanWei}.
\proofend
\end{Rem}

In Appendix \ref{App_2ndES_GiniES}, we point out the fact that the \ $2^{\mathrm{nd}}$-order Expected Shortfall is nothing else but a special Gini Shortfall
 introduced in Furman et al.\ \cite[formula (4.1)]{FurWanZit}, and a (general) Gini Shortfall is a linear combination of
 Expected Shortfall and \ $2^{\mathrm{nd}}$-order Expected Shortfall.
This observation could also underline the importance of studying properties of higher-order (especially, \ $2^{\mathrm{nd}}$-order) Expected Shortfalls.

Next, we recall the notion of probability equivalent level of Value at Risk and Expected Shortfall (abbreviated as $\PELVE$)
 due to Li and Wang \cite[formula (2)]{LiWan}.

\begin{Def}\label{Def_PELVE}
Let \ $X$ \ be a random variable such that \ $X\in L^1$.
The probability equivalent level of Value at Risk and Expected Shortfall
 (abbreviated as \ $\PELVE$) \ of \ $X$ \ at a level \ $\vare\in(0,1)$ \ is defined by
 \begin{align*}
	\Pi_{\vare}(X):=\inf\Big\{c \in \Big[1,\frac{1}{\vare}\Big]: \ES_X(1-c\vare)\leq \VaR_X(1-\vare) \Big\},
 \end{align*}
 where \ $\inf\emptyset=\infty$.
\end{Def}

We give a motivation why the infimum in Definition \ref{Def_PELVE} of \ $\Pi_{\vare}(X)$ \ is taken over \ $\big[1,\frac{1}{\vare}\big]$.
\ The level \ $1-c\vare$ \ of \ $\ES_X$ \ should be non-negative yielding that \ $c\leq \frac{1}{\vare}$; \ and, by \eqref{help_ES2_ES},
 we have \ $\VaR_X(1-c\vare)\leq \ES_X(1-c\vare)$ \ for any \ $c\in(0,\frac{1}{\vare}]$, \
 which together with the requested inequality \ $\ES_X(1-c\vare)\leq \VaR_X(1-\vare)$ \ in the definition of $\Pi_{\vare}(X)$
 imply that \ $\VaR_X(1-c\vare)\leq \VaR_X(1-\vare)$.
\ Provided that \ $[1-\vare,1)\ni p \mapsto \VaR_X(p)$ \ is strictly monotone increasing, this yields that \ $c\geq 1$,
 \ since otherwise \ $1-c\vare > 1-\vare$ \ implying that \ $\VaR_X(1-c\vare)> \VaR_X(1-\vare)$.

Next, we recall a generalization of $\PELVE$ due to Fiori and Rosazza Gianin \cite[Definition 8]{FioGia}.

\begin{Def}\label{Def_Fiori_Gianin}
For each \ $\alpha\in(0,1)$, \ let \ $\varrho_\alpha: L^1\to\RR$ \ be a risk measure such that the family \ $\{\varrho_\alpha: \alpha\in(0,1)\}$ \
 is monotone, i.e., if \ $0<\alpha_1\leq \alpha_2<1$, \ then \ $\varrho_{\alpha_1}(X)\leq \varrho_{\alpha_2}(X)$, \ $X\in L^1$.
For each \ $p\in(0,1)$, \ let us introduce the risk measure \ $\widetilde\varrho_p:L^1\to\RR\cup\{\infty\}$,
 \[
  \widetilde\varrho_p(X):=\frac{1}{1-p}\int_p^1 \varrho_\alpha(X)\,\dd \alpha, \qquad X\in L^1.
 \]
Given a random variable \ $X\in L^1$, \ the generalized $\PELVE$ \ of \ $X$ \ at a level \ $\vare\in(0,1)$ \
 corresponding to the pair \ $( (\varrho_\alpha)_{\alpha\in(0,1)}, (\widetilde\varrho_p)_{p\in(0,1)} )$ \ is defined by
  \begin{align*}
	\Pi_{\vare}^g(X):=\inf\Big\{c \in \Big[1,\frac{1}{\vare}\Big]: \widetilde\varrho_{1-c\vare}(X) \leq \varrho_{1-\vare}(X) \Big\},
 \end{align*}
 where \ $\inf\emptyset=\infty$.
In the special case \ $\varrho_\alpha(X)=\ES_X(\alpha)$, \ $\alpha\in(0,1)$, \ $X\in L^1$, \ the corresponding generalized $\PELVE$ \ is called
 the conditional $\PELVE$ \ (abbreviated as \ c-$\PELVE$).
\end{Def}

By Proposition 7 in Fiori and Rosazza Gianin \cite{FioGia}, the family \ $\{\widetilde\varrho_p: p\in(0,1)\}$ \ is monotone,
 and \ $\widetilde\varrho_p \geq \varrho_p$, \ $p\in(0,1)$.
\ Hence one can give a similar motivation why the infimum in the definition of \ $\Pi_{\vare}^g(X)$ \ is taken over \ $\big[1,\frac{1}{\vare}\big]$ \
 just as we did in case of \ $\Pi_\vare(X)$ \ (see the paragraph after Definition \ref{Def_PELVE}).
Further, note that if one chooses \ $\varrho_\alpha(X)=\VaR_X(\alpha)$, \ $\alpha\in(0,1)$, \ $X\in L^1$ \ in Definition \ref{Def_Fiori_Gianin},
 then the corresponding generalized $\PELVE$ \ is nothing else but (usual) \ $\PELVE$ \ due to Li and Wang \cite{LiWan} recalled in Definition \ref{Def_PELVE}.

Both Li and Wang \cite[Propositions 1-2 and Theorem 1]{LiWan} and Fiori and Rosazza Gianin \cite[Propositions 9-11]{FioGia}
 have studied finiteness, uniqueness, and some properties of \ $\PELVE$ \ and the generalized \ $\PELVE$ \
in Definition \ref{Def_Fiori_Gianin}, respectively.
The \ $\PELVE$ \ values of some notable distributions, such as uniform, exponential, normal, lognormal, \ $t$ \ and Pareto distributions,
 have been calculated or approximated in Li and Wang \cite{LiWan}.
The conditional \ $\PELVE$ \ values of uniform, normal and Pareto distributions have been also calculated in Fiori and Rosazza Gianin \cite{FioGia},
 and it turned out that for uniform and Pareto distributions, the corresponding \ $\PELVE$ \ and conditional \ $\PELVE$ \ values coincide,
 see Fiori and Rosazza Gianin \cite[Subsection 3.2.1]{FioGia}.
Li and Wang \cite[Section 4.2]{LiWan} have described convergence of $\PELVE$ of regularly varying random variables as the level tends to \ $0$,
 \ while Fiori and Rosazza Gianin \cite[Proposition 15]{FioGia} showed that \ $\PELVE$ \ and conditional \ $\PELVE$ \ of a random
 variable having generalized Pareto excess distribution coincide.

In the following definition we replace the Expected Shortfall in Definition \ref{Def_PELVE} by
 the \ $n^{\mathrm{th}}$-order Expected Shortfall, where \ $n\in\NN$.

\begin{Def}\label{Def_PELVE2}
Let \ $X$ \ be a random variable such that \ $X\in L^1$, \ and let \ $n\in\NN$.
\ The probability equivalent level of Value at Risk and \ $n^{\mathrm{th}}$-order Expected Shortfall
 (abbreviated as \ $\PELVE_n$) \ of \ $X$ \ at a level \ $\vare\in(0,1)$ \ is defined by
 \begin{align*}
	\Pi_{\vare,n}(X):=\inf\Big\{c \in \Big[1,\frac{1}{\vare}\Big]: \ES_{X,n}(1-c\vare)\leq \VaR_X(1-\vare) \Big\},
 \end{align*}
 where \ $\inf\emptyset=\infty$.
\end{Def}

First of all, we emphasize that the notion of \ $\PELVE_n$ \ given in Definition \ref{Def_PELVE2}
 is a special case of the so-called distorted \ $\PELVE$ \ introduced in Fiori and Rosazza Gianin \cite[Definition 2]{FioGia}
 that we recall below in Remark \ref{Rem3}.
The research on distorted \ $\PELVE$ \ in Fiori and Rosazza Gianin \cite[Section 3.1]{FioGia}
 and our research on \ $\PELVE_n$ \ have been carried out parallelly, and hence we decided to keep\
 our original presentation as it is, but we always mention those results in Fiori and Rosazza Gianin \cite{FioGia}
 which generalize our results.

\begin{Rem}\label{Rem3}
Let \ $g:[0,1]\to[0,1]$ \ be an increasing and concave function that is continuous at \ $0$ \ with
 \ $g(0)=0$ \ and \ $g(1)=1$.
\ Let us consider the corresponding family of distortions $g_p:[0,1]\to [0,1]$, $p\in[0,1)$, \ given by
 \[
    g_p(x):=\begin{cases}
              g\left(\frac{x}{1-p}\right)  & \text{if \ $x\in[0,1-p)$,}\\
              1  & \text{if \ $x\in[1-p,1)$.}
           \end{cases}
 \]
For each \ $p\in[0,1)$ \ and \ $X\in L^1$, \ let us introduce the so-called Wang premia
 of \ $X$ \ at level \ $p$ \ given by
 \[
    \gVaR_X(p):=\int_0^1 \VaR_X(1-s)\,\dd g_p(s),
 \]
 provided that \ $\int_0^1 \VaR_X(1-s)\,\dd g_p(s)\in\RR$.
Fiori and Rosazza Gianin \cite[Definition 2]{FioGia} have recently introduced the notion of distorted \ $\PELVE$ \
 associated to the family of distortions \ $g_p$, $p\in[0,1)$, \ by replacing \ $\ES_{X,n}(1-c\vare)$ \
 with \ $\gVaR_X(1-c\vare)$ \ in Definition \ref{Def_PELVE2}.
For each \ $n\in\NN$, \ we check that \ $\ES_{X,n}(p)=\gVaR_X(p)$, $p\in[0,1)$, \ with the function
 \ $g(x):=1-(1-x)^n$, \ $x\in[0,1]$, \ which implies that \ $\PELVE_n$ \ is a distorted $\PELVE$ associated to the family of distortions
 \  $g_p$ \ with the given function \ $g$.
\ Namely, with the given function \ $g$, \ we have that
 \begin{align*}
   \gVaR_X(p)& = \int_0^{1-p} \VaR_X(1-s)\,\dd\left(1 - \left(1-\frac{s}{1-p}\right)^n\right)\\
             & = \frac{n}{1-p}\int_0^{1-p} \left(\frac{1-s-p}{1-p}\right)^{n-1}\VaR_X(1-s)\,\dd s
               =  \frac{n}{1-p}\int_p^1 \left(\frac{r-p}{1-p}\right)^{n-1}\VaR_X(r)\,\dd r\\
             & = \ES_{X,n}(p),\qquad p\in[0,1),
 \end{align*}
 as desired.
\proofend
\end{Rem}

Note that, just like the original \ $\PELVE$, \ for each \ $n\in\NN$, \ $\PELVE_n$ \
 is defined under the minimal assumption that the random variable representing the risk has a finite first moment.
Remark also that, since \ $\ES_{X,n}(p)\geq \VaR_X(p)$, \ $p\in[0,1)$ \ (see \eqref{help_ES2_ES}),
 we can give a similar motivation why the infimum in the definition of \ $\Pi_{\vare,n}(X)$ \ is taken over \ $\big[1,\frac{1}{\vare}\big]$ \
 just as we did in case of \ $\Pi_\vare(X)$ \ (see the paragraph after Definition \ref{Def_PELVE}).
Note also that, since \ $\ES_{X,n_1}(p)\leq \ES_{X,n_2}(p)$ \ for \ $p\in[0,1)$ \ and \ $n_1,n_2\in\NN$ \ with \ $n_1\leq n_2$ \
 (see \eqref{help_ES2_ES}), we have $\Pi_{\vare,n_1}(X) \leq \Pi_{\vare,n_2}(X)$ for \ $\vare\in(0,1)$ \ and \ $n_1,n_2\in\NN$ \ with \ $n_1\leq n_2$
 \ (for more details, see Remark \ref{Rem2}).
In particular, \ $\Pi_{\vare}(X) \leq \Pi_{\vare,n}(X)$, $\vare\in(0,1)$, $n\in\NN$.

We will prove results that can be considered as counterparts of the above mentioned results of Li and Wang \cite{LiWan} and Fiori and Rosazza Gianin \cite{FioGia}.
Our forthcoming Propositions \ref{Pro1}, \ref{Pro2} and Theorem \ref{Thm_PELVE_tulajdonsagok} are special cases of
Propositions 3, 4, and 5 in Fiori and Rosazza Gianin \cite{FioGia} for distorted $\PELVE$, respectively.
As we mentioned earlier, the two research works have been carried out parallelly, and hence we decided to present
 proofs of our Propositions \ref{Pro1}, \ref{Pro2} and Theorem \ref{Thm_PELVE_tulajdonsagok}.

The paper is organized as follows.
In Section \ref{Sec_Fin_Uni_Bas_prop} we study the finiteness, uniqueness and some basic properties of \ $\PELVE_n$
 such as inequalities for \ $\PELVE_n$ of sum of comonotonic random variables (see Definition \ref{Def_com_mon}),
 see Propositions \ref{Pro1}, \ref{Pro2} and Theorem \ref{Thm_PELVE_tulajdonsagok}.
In Section \ref{Sec_PELVE2_conv}, under some appropriate conditions, we show that for each \ $n\in\NN$, \ we have
 \ $\Pi_{\vare,n}(X_m)\to \Pi_{\vare,n}(X)$ \ as \ $m\to\infty$ \
 whenever \ $X_m\distr X$ \ as \ $m\to\infty$.
In Section \ref{Sec_PELVE2_notable}, we calculate the \ $\PELVE_n$-values of uniform, exponential and Pareto distributions
 for each $n\geq 2$, \ $n\in\NN$, \ and we approximate the $\PELVE_2$-values of normal distributions.
In particular, it turns out that, for a uniformly distributed random variable \ $X$, \
 the $\PELVE_n$ value $\Pi_{\vare,n}(X)$ of $X$ equals $n+1$ for each $\vare\in(0,\frac{1}{n+1}]$, \ i.e.,
 it is the same constant for $\vare\in(0,\frac{1}{n+1}]$.
\ Similar phenomena occur in case of exponential and Pareto distributions, but not in case of normal distributions.
In Section \ref{Sec_PELVE2_Gen_Par}, we study $\PELVE_2$ of a non-negative random variable having a generalized Pareto excess distribution  function;
 and in Section \ref{Sec_PELVE2_reg_var} we describe the asymptotic behaviour of \ $\PELVE_2$ \ of regularly varying distributions as the level tends to \ $0$.
In Sections \ref{Sec_PELVE2_Gen_Par} and \ref{Sec_PELVE2_reg_var}, we consider
 the \ $\PELVE_2$-values, and not the \ $\PELVE_n$-values of the random variables in question
mainly due to the less complexity of computation in case of \ $\PELVE_2$.
However, note that the \ $2^{\mathrm{nd}}$-order Expected Shortfall plays a central role
 in the decomposition of Gini Shortfall presented in Appendix \ref{App_2ndES_GiniES}.

Section \ref{Sec_sim} is devoted to presenting some simulations and real data analysis for \ $\PELVE_2$ \ on S\&P 500 daily returns.
An interesting phenomenon occurs, \ $\PELVE_2$ \ clearly shows the effect of the COVID-19 pandemic via analysing S\&P 500 daily returns.
We close the paper with three appendices.
In Appendix \ref{App_2ndES_prop}, we study some properties of higher-order Expected Shortfalls given in Definition \ref{Def_ES_n}
 such as finiteness, continuity, monotonicity, additivity for comonotonic random variables and connection with weak convergence.
Our results on finiteness and the connection with weak convergence are in fact consequences of
 recent results of Wang et al.\ \cite[Proposition 1 and Theorem 6]{WanWanWei} on so called distortion risk metrics.
For completeness, we present independent proofs of these results as well.
Appendix \ref{App_2ndES_GiniES} is devoted to develop a connection between \ $2^{\mathrm{nd}}$-order Expected Shortfall and Gini Shortfall.
Finally, in Appendix \ref{Sec_Karamata_zero}, we formulate a Karamata theorem for regularly varying functions at \ $0$ \ with index \ $\kappa>-1$.

\section{Finiteness, uniqueness and basic properties of \ $\PELVE_n$}
\label{Sec_Fin_Uni_Bas_prop}

In what follows, when we write \ $\ES_{X,n}$, \ $\Pi_{\vare,n}(X)$ \ and \ $\PELVE_n$ \ we always mean that \ $n\in\NN$ \
 without mentioning it explicitly.
The following result for \ $\PELVE_n$ \ can be considered as the counterpart of the corresponding result for \ $\PELVE$ \ due to
 Li and Wang \cite[Proposition 1]{LiWan}.
It is a special case of Proposition 3 for distorted $\PELVE$ in Fiori and Rosazza Gianin \cite{FioGia}.

\begin{Pro}\label{Pro1}
Let \ $X$ \ be a random variable such that \ $X\in L^1$, \ $\varepsilon \in(0,1)$ \ and \ $n\in\NN$.
\ Then the following statements are equivalent:
 \begin{itemize}
  \item[(i)] There exists \ $c_0 \in [1,1/\vare]$ \ such that
           \begin{equation}\label{propHO}
	        \ES_{X,n}(1-c_0\vare)=\VaR_X(1-\vare).
	        \end{equation}
  \item[(ii)] $\Pi_{\vare,n}(X) \in [1,1/\vare]$ \ and \eqref{propHO} holds for \ $\Pi_{\vare,n}(X)$, \ i.e.,
              \ $\ES_{X,n}(1-\Pi_{\vare,n}(X)\vare)=\VaR_X(1-\vare)$.
  \item[(iii)] $\ES_{X,n}(0) \leq  \VaR_X(1-\vare)$.
  \item[(iv)]  $\Pi_{\vare,n}(X) < \infty$.
 \end{itemize}
\end{Pro}

\noindent{\bf Proof.}
(i) $\Rightarrow$ (ii):
By (i), the set \ $\{c \in [1,1/\vare]: \ES_{X,n}(1-c\vare)\leq \VaR_X(1-\vare) \}$ \ is nonempty yielding that
 \ $\Pi_{\vare,n}(X) < \infty$ \ and \ $\Pi_{\vare,n}(X) \in [1,1/\vare]$.
By the definition of infimum, there exists a sequence \ $(c_m)_{m \in \NN}$ \ in \ $[1,1/\vare]$ \ such that
 \ $c_m \downarrow \Pi_{\vare,n}(X)$ \ as \ $m\to\infty$, \ and \ $\ES_{X,n}(1-c_m\vare)\leq \VaR_X(1-\vare)$, \ $m\in\NN$.
\ By Lemma \ref{Lem_pelvehofolytonos_monoton}, the function
 \ $[0,1)\ni p \mapsto \ES_{X,n}(p)$ \ is continuous and monotone increasing,
 so it is continuous at the point \ $1-\Pi_{\vare,n}(X)\vare \in [0,1)$, \ and consequently, by taking the limit of both sides of
 \ $\ES_{X,n}(1-c_m\vare)\leq \VaR_X(1-\vare)$ \ as \ $m\to\infty$, \ we have
 \begin{equation}\label{ineq1HO}
  \ES_{X,n}(1-\Pi_{\vare,n}(X)\vare) \leq \VaR_X(1-\vare).
 \end{equation}
Further, using again (i), there exists \ $c_0 \in [1,1/\vare]$ \ such that
 \ $\ES_{X,n}(1-c_0\vare)= \VaR_X(1-\vare)$, and hence, by the definition of infimum, \ $\Pi_{\vare,n}(X) \leq c_0$.
\ Since the function \ $[0,1)\ni p \mapsto \ES_{X,n}(p)$ \ is continuous and monotone increasing,
 \begin{equation}\label {ineq2HO}
   \VaR_X(1-\vare)=\ES_{X,n}(1-c_0\vare) \leq \ES_{X,n}(1-\Pi_{\vare,n}(X)\vare).
 \end{equation}
Inequalities \eqref{ineq1HO} and \eqref{ineq2HO} yield (ii).

(ii) $\Rightarrow$ (iii):
Since the function \ $[0,1)\ni p \mapsto \ES_{X,n}(p)$ \ is continuous and monotone increasing (see Lemma \ref{Lem_pelvehofolytonos_monoton}),
 we have
 \[
   \inf_{c\in [1,1/\vare]}\ES_{X,n}(1-c\vare) = \ES_{X,n}(0) = n \int_{0}^1 s^{n-1} \VaR_X(s) \,\dd s.
 \]
On the contrary to (iii), let us suppose that \ $\ES_{X,n}(0)>\VaR_X(1-\vare)$.
\ Then $\inf_{c\in [1,1/\vare]}\ES_{X,n}(1-c\vare)>\VaR_X(1-\vare)$, and hence
 $\inf\{c \in [1,1/\vare]: \ES_{X,n}(1-c\vare)\leq \VaR_X(1-\vare) \}=\emptyset$.
Consequently, by definition, \ $\Pi_{\vare,n}(X)=\infty$, \ which leads us to a contradiction, since \ $\Pi_{\vare,n}(X)<\infty$ \
 (due to \ $\Pi_{\vare,n}(X)\in[1,1/\vare]$).

(iii) $\Rightarrow$ (iv):  By (iii), choosing \ $c=1/\vare$, \ the set \ $\{c \in [1,1/\vare]: \ES_{X,n}(1-c\vare) \leq \VaR_X(1-\vare) \}$ \ is nonempty,
 so \ $\Pi_{\vare,n}(X) < \infty$, \ as desired.

(iv) $\Rightarrow$ (i):
Using \eqref{help_ES2_ES}, we have \ $\ES_{X,n}(p)\geq \VaR_X(p)$, \ $p\in(0,1)$.
\ In fact, this inequality is a direct consequence of the fact that the function \ $(0,1)\ni p \mapsto \VaR_X(p)$ \ is monotone increasing:
 \begin{align*}
  \ES_{X,n}(p)&=\frac{n}{1-p}\int_{p}^1 \left(\frac{s-p}{1-p}\right)^{n-1} \VaR_X(s) \,\dd s \geq \frac{n}{(1-p)^n} \int_{p}^1 (s-p)^{n-1} \VaR_X(p) \,\dd s\\
    &= \frac{n}{(1-p)^n} \VaR_X(p) \frac{(1-p)^n}{n} = \VaR_X(p).
 \end{align*}
Further, \ $\ES_{X,n}(0)\leq \VaR_X(1-\vare)$, \ since otherwise \ $\Pi_{\vare,n}(X)=\infty$ \ would hold,
 which can be checked similarly as in the proof of part  (ii) $\Rightarrow$ (iii).
Consequently, by the previous two inequalities, we have
 \begin{align}\label{help_ES_3}
  \ES_{X,n}(1-\vare) \geq  \VaR_X(1-\vare) \geq \ES_{X,n}(0).
 \end{align}
Since the function \ $[1,1/\vare]\ni c \mapsto \ES_{X,n}(1-c\vare)$ \ is continuous and monotone decreasing
 (see Lemma \ref{Lem_pelvehofolytonos_monoton}), by Bolzano's intermediate value theorem and \eqref{help_ES_3},
  there exists \ $c_0 \in [1,1/\vare]$ \ such that \eqref{propHO} holds, as desired.
\proofend

In the next remark, we formulate a consequence of part (iii) of Proposition \ref{Pro1},
 and we also compare \ $\Pi_\vare(X)$ \ and \ $\Pi_{\vare,n}(X)$, \ where \ $\vare\in(0,1)$ \ and \ $X\in L^1$.

\begin{Rem}\label{Rem2}
(i). For \ $X\in L^1$ \ and \ $\vare\in(0,1)$, \ the inequality \ $\ES_{X,n}(0) \leq  \VaR_X(1-\vare)$ \ in part (iii) in Proposition \ref{Pro1}
 implies \ $\EE(X) \leq  \VaR_X(1-\vare)$, \ which is nothing else but the inequality in part (iii) in Proposition 1
 in Li and Wang \cite{LiWan}.
Indeed, \ $\EE(X)=\ES_{X}(0)$, \ since
 \[
   \ES_X(0) = \int_0^1 \VaR_X(s)\,\dd s = \EE(\VaR_X(U)),
 \]
 where \ $U$ \ is a uniformly distributed random variable on \ $(0,1)$, \ and the distributions of
 \ $\VaR_X(U)$ \ and \ $X$ \ coincide (see, e.g., Embrechts and Hofert \cite[Proposition 2]{EmbHof}).
Hence, by \eqref{help_ES2_ES}, we have \ $\EE(X)= \ES_X(0)\leq \ES_{X,n}(0)\leq  \VaR_X(1-\vare)$, as desired.

(ii). For \ $X\in L^1$, \ the inequality \ $\ES_{X,n_1}(p)\leq \ES_{X,n_2}(p)$ \ for \ $p\in[0,1)$ \ and \ $n_1,n_2\in\NN$ \ with \ $n_1\leq n_2$ \
 (see \eqref{help_ES2_ES}) yields \ $\Pi_{\vare,n_1}(X)\leq \Pi_{\vare,n_2}(X)$ \ for \ $\vare\in(0,1)$ \ and \ $n_1,n_2\in\NN$ \ with \ $n_1\leq n_2$.
In particular, \ $\Pi_{\vare}(X) \leq \Pi_{\vare,n}(X)$ \ for \ $\vare\in(0,1)$ \ and \ $n\in\NN$.
\proofend
\end{Rem}

The following result for \ $\PELVE_n$ \ can be considered as the counterpart of the corresponding result for \ $\PELVE$ \ due to
 Li and Wang \cite[Proposition 2]{LiWan}.
It is a special case of Proposition 4 for distorted $\PELVE$ in Fiori and Rosazza Gianin \cite{FioGia}.

\begin{Pro}\label{Pro2}
Let \ $X$ \ be a random variable such that \ $X\in L^1$, \ $\varepsilon \in(0,1)$, \ and \ $n\in\NN$.
\ Let us suppose that the function \ $(0,1)\ni p \mapsto \VaR_X(p)$ \
 is not constant on the interval \ $[1-\vare,1)$, and let us assume that \ $\ES_{X,n}(0) \leq  \VaR_X(1-\vare)$.
\ Then there exists a unique \ $c_0 \in [1, 1/\vare]$ \ such that \eqref{propHO} holds.
\end{Pro}

\noindent{\bf Proof.}
By part (iii) $\Rightarrow$ (i) of Proposition \ref{Pro1}, there exists \ $c_0 \in [1, 1/\vare]$ \ such that \eqref{propHO} holds.
Further, the function \ $[0,1)\ni p \mapsto \ES_{X,n}(p)$ \ is continuous and it is strictly monotone increasing on \ $[0,1-\vare]$ \
 (see Lemmas \ref{Lem_pelvehofolytonos_monoton} and \ref{Lem_szigmonHO}), yielding that the function \ $[1, 1/\vare] \ni c \mapsto \ES_{X,n}(1-c\vare)$ \
 is continuous and strictly monotone decreasing.
This together with the existence of \ $c_0\in [1, 1/\vare]$ \ satisfying \eqref{propHO} yield the uniqueness of such a \ $c_0$, \ as desired.
\proofend

Note that the assumption \ $\ES_{X,n}(0) \leq  \VaR_X(1-\vare)$ \ in Proposition \ref{Pro2} yields that
 \ $\ES_X(0) \leq  \VaR_X(1-\vare)$ \ (due to the second inequality in \eqref{help_ES2_ES}),
 which is nothing else but the corresponding condition for \ $\PELVE$ \ in Proposition 2 in Li and Wang \cite{LiWan}.
Consequently, under the assumptions of Proposition \ref{Pro2} we also have that
 \ $\Pi_{\vare}(X)<\infty$ \ and \ $\ES_X(1-\Pi_\vare(X)\vare)=\VaR_X(1-\vare)$.

The following result for \ $\PELVE_n$ \ can be considered as the counterpart of the corresponding result for \ $\PELVE$ \ due to
 Li and Wang \cite[Theorem 1]{LiWan}.
It is a special case of Proposition 5 for distorted $\PELVE$ in Fiori and Rosazza Gianin \cite{FioGia}.

\begin{Thm}\label{Thm_PELVE_tulajdonsagok}
Let \ $X$ \ be a random variable such that \ $X\in L^1$, \ $\varepsilon \in(0,1)$, \ and \ $n\in\NN$.
\ Let us suppose that \ $\ES_{X,n}(0)\leq \VaR_X(1-\vare)$ \ holds.
Then the following statements hold:
 \begin{itemize}
  \item[(i)] scale-location invariance: \ $\Pi_{\vare,n}(\lambda X + a) = \Pi_{\vare,n}(X)$ \ for each \ $\lambda>0$ \ and \ $a\in\RR$.
  \item[(ii)] $\Pi_{\vare,n}(f(X)) \leq \Pi_{\vare,n}(X)$ \ for each monotone increasing and concave function \ $f:\RR\to\RR$ \ with \ $f(X)\in L^1$.
  \item[(iii)] $\Pi_{\vare,n}(X) \leq \Pi_{\vare,n}(g(X))$ \ for each strictly monotone increasing and convex function \ $g:\RR\to\RR$ \ with
              \ $\range(g)=\RR$ \ and \ $g(X)\in L^1$.
  \item[(iv)] quasi-convexity and quasi-concavity for comonotonic random variables:
              \begin{align*}
                \min\{ \Pi_{\vare,n}(f(X)), \Pi_{\vare,n}(g(X)) \}
                    &\leq \Pi_{\vare,n}(\lambda f(X) + (1-\lambda) g(X))\\
                    &\leq \max\{ \Pi_{\vare,n}(f(X)), \Pi_{\vare,n}(g(X)) \}
              \end{align*}
              for each $\lambda\in[0,1]$ and monotone increasing functions $f,g:\RR\to\RR$ with $f(X),g(X)\in L^1$.
 \end{itemize}
\end{Thm}

\noindent{\bf Proof.}
First, we prove that (ii) yields (i).
Let\ $\lambda>0$, \ $a\in\RR$, \ $f:\RR\to\RR$, \ $f(x):=\lambda x+ a$, \ $x\in\RR$, \ and \ $g:\RR\to\RR$, \ $g(y):=\frac{1}{\lambda}(y-a)$, \ $y\in\RR$.
\ Then \ $f$ \ and \ $g$ \ are monotone increasing, linear (hence convex and concave) functions, and they are inverses of each other.
Further, \ $f(X),g(X)\in L^1$, \ since \ $\EE(\vert f(X)\vert)\leq \lambda \EE(\vert X\vert) + \vert a\vert<\infty$ \ and
 \ $\EE(\vert g(X)\vert)\leq \frac{1}{\lambda} \EE(\vert X\vert) + \frac{\vert a\vert}{\lambda}<\infty$ \ due to \ $X\in L^1$.
\ Consequently, by (ii),
 \[
   \Pi_{\vare,n}(f(X)) \leq \Pi_{\vare,n}(X)= \Pi_{\vare,n}((g\circ f)(X)) \leq \Pi_{\vare,n}(f(X)),
 \]
 yielding \ $\Pi_{\vare,n}(f(X)) = \Pi_{\vare,n}(X)$, \ i.e., (i), as desired.

Now we prove (ii).
Let \ $f:\RR\to\RR$ \ be a monotone increasing and concave function with \ $f(X)\in L^1$. \
Since \ $f$ \ is concave and defined on (the open interval) \ $\RR$, \ we have \ $f$ \ is continuous.
Further, it is known that for any monotone increasing and continuous function \ $h:\RR\to\RR$, \ we have
 \ $\VaR_{h(X)}(p) = h(\VaR_X(p))$, \ $p\in(0,1)$, \ see, e.g., Shorack and Wellner \cite[Exercise 3, page 9]{ShoWel}
  or Dhaene et al.\ \cite[part (a) of Theorem 1]{DhaDenGooKaaVyn}.
Consequently, we get
 \begin{align}\label{help_PELVE_1}
 \VaR_{f(X)}(p) = f(\VaR_X(p)), \qquad p\in(0,1).
 \end{align}
Let \ $c_1:=\Pi_{\vare,n}(X)$ \ and \ $c_2:=\Pi_{\vare,n}(f(X))$.
\ Since, by assumption, \ $\ES_{X,n}(0)\leq \VaR_X(1-\vare)$, \ using the equivalence of (ii), (iii) and (iv) in Proposition \ref{Pro1}, we have
 \ $c_1<\infty$ \ and
 \begin{align}\label{help_PELVE_2}
   \ES_{X,n}(1-c_1\vare)=\VaR_X(1-\vare).
 \end{align}
Note that for each \ $p\in[0,1)$, \ the function \ $\sigma_p:[0,1)\to[0,\infty)$, \ $\sigma_p(s):=\frac{n}{(1-p)^n}(s-p)^{n-1}\bone_{[p,1]}(s)$, \ $s\in[0,1]$, \
 is a distortion function in the sense of Definition 3.6 in Pflug and Pichler \cite{PflPic}, since it is non-negative, monotone increasing and
 \ $\int_0^1\sigma_p(s)\,\dd s=1$.
\ Hence for each \ $p\in[0,1)$ \ one can apply Corollary 3.19 in Pflug and Pichler \cite{PflPic} with the distortion function \ $\sigma_p$, and we have
 \[
     \ES_{f(X),n}(p) = \sup_{\text{$U$ \ is uniformly distributed on \ $[0,1]$}} \EE(f(X)\sigma_p(U)), \qquad p\in[0,1).
 \]
Here we implicitly assumed that the underlying probability space is rich enough such that there exists
 a random variable on it with uniform distribution on \ $[0,1]$.
\ This assumption is not a restriction, see, e.g., Rachev \cite[Theorem 2.5.1]{Rac}.
In particular, if the underlying probability space \ $(\Omega,\cF,\PP)$ \ is atomless,
 then there exists a  random variable \ $U:\Omega\to\RR$ \ with uniform distribution on \ $[0,1]$,
 \ see, e.g., Rachev \cite[Theorems 2.4.1 and 2.5.2]{Rac}.
Further, one can check that for any random variable \ $U$ \ which is uniformly distributed on \ $[0,1]$, \ we have
 \ $\QQ(A):=\int_A (\sigma_p(U))(\omega)\PP(\dd\omega)$, \ $A\in\cF$, \ is a probability measure
 on \ $(\Omega,\cF)$ \ such that \ $\QQ$ \ is absolutely continuous with respect to \ $\PP$ \ and
 \ $\EE(f(X)\sigma_p(U)) = \EE_\QQ(f(X))$.
\ Hence, for any random variable \ $U$ \ which is uniformly distributed on \ $[0,1]$, \ by Jensen's inequality, we get
 \[
    \EE(f(X)\sigma_p(U)) = \EE_\QQ(f(X)) \leq f(\EE_\QQ(X)) = f(\EE(X\sigma_p (U))).
 \]
Consequently, using also that \ $f$ \ is monotone increasing, we have
 \begin{align}\label{help_PELVE_3}
 \begin{split}
  \ES_{f(X),n}(p) & \leq \sup_{\text{$U$ \ is uniformly distributed on \ $[0,1]$}} f(\EE(X\sigma_p (U)))\\
                  & \leq f\Big(\sup_{\text{$U$ \ is uniformly distributed on \ $[0,1]$}}\EE(X\sigma_p (U)) \Big)\\
                  & = f(\ES_{X,n}(p)),  \qquad p\in[0,1).
 \end{split}
 \end{align}
Since the functions \ $f$, \ $[0,1)\ni p\mapsto \ES_{X,n}(p)$ \ and \ $[0,1)\ni p\mapsto \ES_{f(X),n}(p)$ \ are continuous
 (see Lemma \ref{Lem_pelvehofolytonos_monoton}), by taking the limit of both sides of the inequality \eqref{help_PELVE_3} as \ $p\downarrow 0$, \ we have
 \[
   \ES_{f(X),n}(0) \leq f(\ES_{X,n}(0)) \leq f(\VaR_X(1-\vare)) = \VaR_{f(X)}(1-\vare),
 \]
 where, for the second inequality, we used that \ $f$ \ is monotone increasing and \ $\ES_{X,n}(0)\leq \VaR_X(1-\vare)$ \ (by assumption),
 and, for the equality, \eqref{help_PELVE_1}.
So, by Proposition \ref{Pro1}, we have \ $c_2=\Pi_{\vare,n}(f(X))<\infty$ \ and \ $\ES_{f(X),n}(1-c_2\vare)=\VaR_{f(X)}(1-\vare)$.
\ Using \eqref{help_PELVE_3} with the choice of \ $p:=1-c_1\vare$, \ \eqref{help_PELVE_2} and \eqref{help_PELVE_1}, we get
 \[
   \ES_{f(X),n}(1-c_1\vare) \leq f(\ES_{X,n}(1-c_1\vare)) = f(\VaR_X(1-\vare)) = \VaR_{f(X)}(1-\vare),
 \]
 and, by Definition \ref{Def_PELVE2} of $\PELVE_n$, we have $\Pi_{\vare,n}(f(X))\leq c_1$, i.e.,
 $\Pi_{\vare,n}(f(X))\leq \Pi_{\vare,n}(X)$, as desired.

(iii).
Since \ $g$ \ is a convex function defined on \ $\RR$, \ it is continuous.
Further, due to our assumptions, \ $g^{-1}:\RR\to \RR$ \ is a strictly monotone increasing and concave function with \ $\range(g^{-1})=\RR$,
 and since \ $g(X)\in L^1$ \ and \ $g^{-1}(g(X))=X\in L^1$, \ part (ii) yields that
 \begin{align*}
   \Pi_{\vare,n}(X) = \Pi_{\vare,n}( g^{-1}(g(X)) )  \leq \Pi_{\vare,n}(g(X)),
 \end{align*}
 as desired.

(iv).
First, we check that for any \ $c,d\in[1,\frac{1}{\vare}]$ \ and \ $Y\in L^1$, \ we have
 \begin{align}\label{help_PELVE_4}
  \begin{split}
  &d < \Pi_{\vare,n}(Y)\leq c\\
  &\qquad \Longleftrightarrow\qquad \ES_{Y,n}(1-c\vare)\leq \ES_{Y,n}(1-\Pi_{\vare,n}(Y)\vare)=\VaR_Y(1-\vare)<\ES_{Y,n}(1-d\vare).
 \end{split}
 \end{align}

Proof of part \ $\Longrightarrow$ \ of \eqref{help_PELVE_4}:
If \ $\Pi_{\vare,n}(Y)< c$, \ then, by the definition of infimum, there exists \ $a<c$ \ such that \ $a\in[1,\frac{1}{\vare}]$ \ and
 \ $\ES_{Y,n}(1-a\vare) \leq \VaR_Y(1-\vare)$.
\ Since the function \ $[0,1)\ni p \mapsto \ES_{Y,n}(p)$ \ is monotone increasing (see Lemma \ref{Lem_pelvehofolytonos_monoton}),
 we have
 \[
   \ES_{Y,n}(1-c\vare) \leq \ES_{Y,n}(1-a\vare) \leq  \VaR_Y(1-\vare).
 \]
If \ $\Pi_{\vare,n}(Y) = c$, \ then, again by the definition of infimum, there exists a sequence \ $(c_m)_{m\in\NN}$ \ in \ $[1,\frac{1}{\vare}]$ \
 such that \ $c_m\downarrow c$ \ as \ $m\to\infty$ \ and
 \[
   \ES_{Y,n}(1-c_m\vare) \leq \VaR_Y(1-\vare),\qquad m\in\NN.
 \]
Using that the function \ $[0,1)\ni p\mapsto \ES_{Y,n}(p)$ \ is continuous (see Lemma \ref{Lem_pelvehofolytonos_monoton}),
 by taking the limit of both sides of the inequality above as \ $m\to\infty$, \ we have \ $\ES_{Y,n}(1-c\vare) \leq \VaR_Y(1-\vare)$.
\ Further, since
 \[
   d< \Pi_{\vare,n}(Y) = \inf\Big\{a\in\Big[1,\frac{1}{\vare}\Big] : \ES_{Y,n}(1-a\vare)\leq \VaR_Y(1-\vare) \Big\},
 \]
 we have \ $\ES_{Y,n}(1-d\vare)>\VaR_Y(1-\vare)$, \ as desired.

Proof of part \ $\Longleftarrow$ \ of \eqref{help_PELVE_4}:
Since \ $\ES_{Y,n}(1-c\vare) \leq \VaR_Y(1-\vare)$, \ we have
 \[
   c\in \Big\{a\in\Big[1,\frac{1}{\vare}\Big] : \ES_{Y,n}(1-a\vare)\leq \VaR_Y(1-\vare) \Big\},
 \]
 and, by the definition of \ $\Pi_{\vare,n}(Y)$, \ we get \ $\Pi_{\vare,n}(Y) \leq c$.
\ Next we check that \ $d>\Pi_{\vare,n}(Y)$ \ or \ $d=\Pi_{\vare,n}(Y)$ \ cannot hold,
 so \ $d<\Pi_{\vare,n}(Y)$, \ as desired.
If \ $d>\Pi_{\vare,n}(Y)$ \ would hold, then, by the definition of infimum, there exists \ $\widehat d\in[1,\frac{1}{\vare}]$ \ such that
 \ $\widehat d < d$ \ and \ $\ES_{Y,n}(1-\widehat d\vare)\leq \VaR_Y(1-\vare)$.
\ Since the function \ $[0,1)\ni p\mapsto \ES_{Y,n}(p)$ \ is monotone increasing, we have
 \ $\ES_{Y,n}(1-d\vare)\leq \ES_{Y,n}(1-\widehat d\vare)$, \ and hence
 \ $\ES_{Y,n}(1-d\vare)\leq \VaR_Y(1-\vare)$.
\ This leads us to a contradiction, since, by assumption, \ $\VaR_Y(1-\vare) < \ES_{Y,n}(1-d\vare)$,
 \ and consequently, \ $d>\Pi_{\vare,n}(Y)$ \ cannot hold.
If \ $d=\Pi_{\vare,n}(Y)$ \ would hold, then similarly as in proving  part \ $\Longrightarrow$ \ of \eqref{help_PELVE_4},
 by definition of infimum and the continuity of the function \ $[0,1)\ni p\mapsto \ES_{Y,n}(p)$, \ we have
 \ $\ES_{Y,n}(1-d\vare) \leq \VaR_Y(1-\vare)$. \
This leads us to a contradiction, since, by assumption, \ $\VaR_Y(1-\vare) < \ES_{Y,n}(1-d\vare)$, \ and consequently
 \ $d=\Pi_{\vare,n}(Y)$ \ cannot hold.

Since \ $f,g:\RR\to\RR$ \ are monotone increasing and \ $\lambda\in[0,1]$, \ we have
 \ $\lambda f(X)$ \ and \ $(1-\lambda) g(X)$ \ are comonotonic random variables,
 so, by the additivity of \ $\VaR$ \ for comonotonic random variables (see, e.g., McNeil et al.\ \cite[Proposition 7.20]{McnFreEmb}
 or Dhaene et al.\ \cite[Theorem 4.2.1]{DhaVanGooKaaTanVyn}), we get
 \[
  \VaR_{\lambda f(X) + (1-\lambda)g(X)}(1-\vare)
     =\VaR_{\lambda f(X)}(1-\vare) + \VaR_{(1-\lambda) g(X)}(1-\vare).
 \]
Let \ $c_1:=\Pi_{\vare,n}(f(X))$ \ and \ $c_2:=\Pi_{\vare,n}(g(X))$.
\ Let us suppose that \ $c_1=\infty$ \ and \ $c_2=\infty$.
\ In this case \ $\min(c_1,c_2)=\infty$ \ and \ $\max(c_1,c_2)=\infty$, \ so it is enough to check that
 \ $\Pi_{\vare,n}(\lambda f(X) + (1-\lambda)g(X))=\infty$.
\ By Proposition \ref{Pro1}, the positive homogeneity and comonotonic additivity of Value at Risk and $n^{\mathrm{th}}$-order Expected Shortfall
 (see Proposition \ref{Pro_ESn_com_add}), we have
 \begin{align*}
 &\Pi_{\vare,n}(\lambda f(X) + (1-\lambda)g(X))=\infty
   \quad \Leftrightarrow \quad
 \ES_{\lambda f(X) + (1-\lambda)g(X), n}(0) > \VaR_{\lambda f(X) + (1-\lambda)g(X)}(1-\vare)\\
   &  \quad \Leftrightarrow \quad
  \lambda  \ES_{f(X),n}(0) + (1-\lambda) \ES_{g(X),n}(0)  > \lambda \VaR_{f(X)}(1-\vare) + (1-\lambda) \VaR_{g(X)}(1-\vare).
 \end{align*}
Here the last inequality is satisfied, since using again Proposition \ref{Pro1}, \ $c_1=\Pi_{\vare,n}(f(X))=\infty$, \ $c_2=\Pi_{\vare,n}(f(X))=\infty$, \
 and \ $f(X), g(X)\in L^1$, \ we have
 \[
    \ES_{f(X),n}(0) > \VaR_{f(X)}(1-\vare)  \qquad \text{and} \qquad \ES_{g(X),n}(0) > \VaR_{g(X)}(1-\vare).
 \]

Let us suppose now that at least one of \ $c_1$ \ and \ $c_2$ \ is finite.
Then for each \ $d<\min\{c_1,c_2\}$ \ with \ $d\in[1,\frac{1}{\vare}]$, \ using \eqref{help_PELVE_4} and Proposition \ref{Pro1},
 we get
 \[
   \VaR_{f(X)}(1-\vare) < \ES_{f(X),n}(1-d\vare) \qquad \text{and}\qquad  \VaR_{g(X)}(1-\vare) < \ES_{g(X),n}(1-d\vare).
 \]
Indeed, if both \ $c_1$ \ and \ $c_2$ \ are finite, then it readily follows by \eqref{help_PELVE_4};
 and if \ $c_1<\infty$ \ and \ $c_2=\infty$, \ then, by \eqref{help_PELVE_4},
 \ $\VaR_{f(X)}(1-\vare) < \ES_{f(X),n}(1-d\vare)$, \ and, by Proposition \ref{Pro1} and the monotone increasing
 property of the function \ $[0,1)\ni p\mapsto \ES_{g(X),n}(p)$, \ we have \ $\VaR_{g(X)}(1-\vare) < \ES_{g(X),n}(0) \leq \ES_{g(X),n}(1-d\vare)$,
 \ as desired.
The case \ $c_1=\infty$ \ and \ $c_2<\infty$ \ can be handled similarly.

\noindent As a consequence, for each \ $\lambda\in[0,1]$, \ we have
 \[
    \lambda \VaR_{f(X)}(1-\vare) + (1-\lambda)\VaR_{g(X)}(1-\vare) < \lambda \ES_{f(X),n}(1-d\vare) + (1-\lambda)  \ES_{g(X),n}(1-d\vare),
 \]
 and using again the positive homogeneity and comonotonic additivity of Value at Risk and \ $n^{\mathrm{th}}$-order Expected Shortfall, we have
 \[
     \VaR_{\lambda f(X)+ (1-\lambda)g(X)}(1-\vare)  <  \ES_{\lambda f(X) + (1-\lambda) g(X),n}(1-d\vare),\qquad \lambda\in[0,1].
 \]
Hence, by Definition \ref{Def_PELVE2} and the continuity and monotone increasing property of the function
 \ $[0,1)\ni p \mapsto \ES_{\lambda f(X) + (1-\lambda) g(X),n}(p)$, \ we have
  \ $d< \Pi_{\vare,n}(\lambda f(X) + (1-\lambda)g(X))$ \ for \ $\lambda\in[0,1]$ \ and \ $d<\min\{c_1,c_2\}$.
\ By taking the limit \ $d\uparrow \min\{c_1,c_2\}$, \ we have
 \[
   \min\{c_1,c_2\} \leq \Pi_{\vare,n}(\lambda f(X) + (1-\lambda)g(X)),
 \]
 as desired.

If \ $\max(c_1,c_2)=\infty$, \ then the second inequality in (iv) automatically holds.
If \ $\max(c_1,c_2)<\infty$, \ i.e., both \ $c_1$ \ and \ $c_2$ \ are finite, then let \ $c:=\max\{c_1,c_2\}$.
\ Then \ $\Pi_{\vare,n}(f(X)) = c_1\leq c$ \ and \ $\Pi_{\vare,n}(g(X)) = c_2\leq c$, \ so, by \eqref{help_PELVE_4},
 \[
    \ES_{f(X),n}(1-c\vare) \leq  \VaR_{f(X)}(1-\vare)
     \qquad \text{and}\qquad
    \ES_{g(X),n}(1-c\vare) \leq  \VaR_{g(X)}(1-\vare).
 \]
Hence for each \ $\lambda\in[0,1]$,
 \[
   \lambda \ES_{f(X),n}(1-c\vare) + (1-\lambda) \ES_{g(X),n}(1-c\vare) \leq \lambda \VaR_{f(X)}(1-\vare) + (1- \lambda) \VaR_{g(X)}(1-\vare),
 \]
 and then the positive homogeneity and comonotonic additivity of \ $\VaR$ \ and \ $n^{\mathrm{th}}$-order Expected Shortfall
 yield that
 \[
     \ES_{\lambda f(X) +  (1-\lambda) g(X),n}(1-c\vare) \leq \VaR_{\lambda f(X)+  (1-\lambda) g(X) }(1-\vare).
 \]
Consequently, using  \eqref{help_PELVE_4} with \ $Y:=\lambda f(X) + (1-\lambda)g(X)$, \ we have
 \[
   \Pi_{\vare,n}(\lambda f(X) +  (1-\lambda) g(X))\leq c = \max\{c_1,c_2\},
 \]
 as desired.
\proofend

In the next remark, we point out that part (iii) of Theorem \ref{Thm_PELVE_tulajdonsagok} does not hold for a general monotone
 increasing and convex function \ $g:\RR\to\RR$.

\begin{Rem}
Let \ $g:\RR\to\RR$, \ $g(x)=A$, \ $x\in\RR$, \ with some \ $A\in\RR$.
\ Then  for any random variable \ $X$, \ $n\in\NN$, \ and \ $p\in(0,1)$, \ we have
 \ $\VaR_{g(X)}(p)=A$ \ and
 \[
    \ES_{g(X),n}(p) = \frac{n}{1-p}\int_p^1 \left(\frac{s-p}{1-p}\right)^{n-1}\VaR_{g(X)}(s)\,\dd s
                   = \frac{nA}{(1-p)^n} \int_p^1 (s-p)^{n-1}\,\dd s
                   = A,
 \]
 and hence for each \ $\vare\in(0,1)$,
 \[
  \Pi_{\vare,n}(g(X)) = \inf\Big\{c \in \Big[1,\frac{1}{\vare}\Big]: \ES_{g(X),n}(1-c\vare)\leq \VaR_{g(X)}(1-\vare) \Big\}
                        =\inf\Big[1,\frac{1}{\vare}\Big]
                        =1.
 \]
Consequently, if $X$ is random variable such that $\Pi_{\vare,n}(X)>1$, then $\Pi_{\vare,n}(g(X))\geq \Pi_{\vare,n}(X)$ cannot hold.
Note that $g$ is not strictly increasing and $\range(g) = \{A\}\ne \RR$.
All in all, part (iii) of Theorem \ref{Thm_PELVE_tulajdonsagok} does not hold for a general monotone
 increasing and convex function $g:\RR\to\RR$.
\proofend
\end{Rem}

\section{Convergence properties of \ $\PELVE_n$}\label{Sec_PELVE2_conv}

The following result for \ $\PELVE_n$ \ can be considered as the counterpart of the corresponding result for \ $\PELVE$ \ due to Li and Wang \cite[Theorem 2]{LiWan}.

\begin{Thm}\label{Thm_PELVE2_conv}
Let \ $X_m\in L^1$, \ $m\in\NN$, \ and \ $X\in L^1$ \ be random variables.
Let \ $\vare\in(0,1)$ \ and \ $n\in\NN$.
\ If
 \begin{itemize}
  \item[(i)] $\ES_{X,n}(0)< \VaR_X(1-\vare)$,
  \item[(ii)] the function \ $(0,1)\ni p \mapsto \VaR_X(p)$ \ is not constant on the interval \ $[1-\vare,1)$ and
              it is continuous at \ $1-\vare$,
  \item[(iii)] $X_m\distr X$ \ as \ $m\to\infty$,
  \item[(iv)] $\{X_m : m\in\NN\}$ \ is uniformly integrable,
 \end{itemize}
 then \ $\Pi_{\vare,n}(X_m)\to \Pi_{\vare,n}(X)$ \ as \ $m\to\infty$.
\end{Thm}

\noindent{\bf Proof.}
By the second part of (ii) and (iii), the quantile convergence theorem (see, e.g., Shorack and Wellner \cite[Exercise 5, page 10]{ShoWel})
 yields that
 \begin{align}\label{help_PELVE_5}
   \VaR_{X_m}(1-\vare)\to \VaR_X(1-\vare) \qquad \text{ as \ $m\to\infty$.}
 \end{align}
Using (iii), (iv) and Lemma \ref{Lem_ES2_conv}, we have
  \begin{align}\label{help_PELVE_6}
  \ES_{X_m,n}(t)\to \ES_{X,n}(t)\qquad \text{as \ $m\to\infty$ \ \ for each \ $t\in[0,1)$.}
 \end{align}

Let us introduce the functions \ $f_m:[0,1)\to\RR$, \ $m\in\NN$, \ and \ $f:[0,1)\to\RR$, \ given by
 \[
   f_m(t):= \ES_{X_m,n}(t) - \VaR_{X_m}(1-\vare),\qquad t\in[0,1),
 \]
 and
 \[
   f(t):= \ES_{X,n}(t) - \VaR_X(1-\vare),\qquad t\in[0,1).
 \]
By \eqref{help_PELVE_5} and \eqref{help_PELVE_6}, we have \ $f_m$ \ converges pointwise to \ $f$ \ on \ $[0,1)$ \ as \ $m\to\infty$, \
 and, using also Lemma \ref{Lem_pelvehofolytonos_monoton}, we get that \ $f_m$, \ $m\in\NN$, \ and \ $f$ \ are continuous and monotone increasing functions.
Further, the first part of (ii) and Lemma \ref{Lem_szigmonHO} yield that \ $f$ \ is strictly monotone increasing on the interval \ $[0,1-\vare]$.

Let us recall the following result from calculus: given \ $a<b$, \ $a,b\in\RR$, \ and a sequence of monotone increasing
 real-valued functions on \ $[a,b]$ \ converging pointwise to a continuous function, we have that the convergence holds uniformly on \ $[a,b]$ \ as well.

The above recalled result yields that \ $f_m$ \ converges uniformly on any interval \ $[0,1-\delta]$ \ to \ $f$ \ as \ $m\to\infty$, \ where \ $\delta\in(0,1)$.

Let us consider the reparametrizations \ $g_m:[1,\frac{1}{\vare}]\to\RR$, \ $m\in\NN$, \ and
 \ $g:[1,\frac{1}{\vare}]\to\RR$ \ of \ $f_m$, \ $m\in\NN$, \ and \ $f$, \ respectively, given by
 \[
  g_m(c):=f_m(1-c\vare)= \ES_{X_m,n}(1-c\vare) - \VaR_{X_m}(1-\vare), \qquad c\in \left[1,\frac{1}{\vare}\right],
 \]
 and
 \[
  g(c):=f(1-c\vare)= \ES_{X,n}(1-c\vare) - \VaR_X(1-\vare), \qquad c\in \left[1,\frac{1}{\vare}\right].
 \]
Then \ $g_m$, \ $m\in\NN$, \ and \ $g$ \ are continuous and monotone decreasing functions, and \ $g$ \ is strictly monotone decreasing as well.
Further, $g_m$ converges uniformly on $[1,\frac{1}{\vare}]$ to $g$ as $m\to\infty$.

Using (i), \eqref{help_PELVE_5} and \eqref{help_PELVE_6} with \ $t=0$, \ we have \ $\ES_{X_m,n}(0)< \VaR_{X_m}(1-\vare)$ \ for large enough \ $m\in\NN$.
\ Hence, by Proposition \ref{Pro1}, for large enough \ $m\in\NN$, \ we have \ $\Pi_{\vare,n}(X_m)\in [1,\frac{1}{\vare}]$ \ and
 \ $\Pi_{\vare,n}(X_m)$ \ solves the equation \ $g_m(c)=0$, \ $c\in[1,\frac{1}{\vare}]$, \ and we also have
 \ $\Pi_{\vare,n}(X)\in [1,\frac{1}{\vare}]$ \ and \ $\Pi_{\vare,n}(X)$ \ solves the equation
 \ $g(c)=0$, \ $c\in[1,\frac{1}{\vare}]$.

Then for each \ $m\in\NN$,
 \begin{align*}
 \vert g(\Pi_{\vare,n}(X_m))\vert
  = \vert g(\Pi_{\vare,n}(X_m)) - g_m(\Pi_{\vare,n}(X_m))\vert
  \leq \sup_{c\in[1,\frac{1}{\vare}]}\vert g_m(c) - g(c)\vert
  \to 0 \qquad \text{as \ $m\to\infty$,}
 \end{align*}
 since \ $g_m$ \ converges uniformly on \ $[1,\frac{1}{\vare}]$ \ to \ $g$ \ as \ $m\to\infty$.
\ So \ $\lim_{m\to\infty} g(\Pi_{\vare,n}(X_m))=0$.
\ Further, if \ $\widetilde c\in[1,\frac{1}{\vare}]$ \ is a limit point of the sequence \ $(\Pi_{\vare,n}(X_m))_{m\in\NN}$, \ then there
 exists a subsequence \ $(\Pi_{\vare,n}(X_{m_k}))_{k\in\NN}$ \ in \ $[1,\frac{1}{\vare}]$ \ such that \ $\Pi_{\vare,n}(X_{m_k})\ne \widetilde c$, \ $k\in\NN$, \
 and \ $\Pi_{\vare,n}(X_{m_k})\to \widetilde c$ \ as \ $k\to\infty$.
Since \ $g$ \ is continuous, we have \ $g(\Pi_{\vare,n}(X_{m_k}))\to g(\widetilde c)$ \ as \ $k\to\infty$, \ where
 \ $g(\widetilde c)=0$ \ due to \ $\lim_{m\to\infty} g(\Pi_{\vare,n}(X_m))=0$.
\ Hence \ $\widetilde c\in[1,\frac{1}{\vare}]$ \ is a root of \ $g$, \ and using that \ $g$ \ has a unique root \ $\Pi_{\vare,n}(X)$ \ on \ $[1,\frac{1}{\vare}]$ \
 (since we already checked that \ $\Pi_{\vare,n}(X)$ \ is a root of \ $g$ \ and \ $g$ \ is continuous and strictly monotone decreasing), we get \ $\widetilde c = \Pi_{\vare,n}(X)$.
\ All in all, for any limit point \ $\widetilde c$ \ of \ $(\Pi_{\vare,n}(X_m))_{m\in\NN}$, \ we have \ $\widetilde c = \Pi_{\vare,n}(X)$.
\ Since \ $(\Pi_{\vare,n}(X_m))_{m\in\NN}$ \ is a bounded sequence in \ $[1,\frac{1}{\vare}]$, \ it has a limit point,
 and taking into account our previous considerations, \ $\Pi_{\vare,n}(X_m)$ \ converges to its unique limit point \ $\Pi_{\vare,n}(X)$ \ as \ $m\to\infty$.
\proofend

\section{$\PELVE_n$ \ of some notable distributions}\label{Sec_PELVE2_notable}

In this section, we calculate the \ $\PELVE_n$-values of uniform, exponential and Pareto distributions
 for each $n\geq 2$, \ $n\in\NN$, \ and we approximate the $\PELVE_2$-values of normal distributions.

\begin{Ex}[Uniform distribution]
Let \ $X$ \ be a random variable with uniform distribution on the interval \ $[0,1]$, \ and let \ $n\in\NN$.
\ Then \ $\VaR_X(p) = p$, \ $p\in(0,1)$, \ and
\begin{align*}
 \ES_{X,n}(p)
  & = \frac{n}{(1-p)^n} \int_p^1 (s-p)^{n-1}\VaR_X(s)\,\dd s\\
  & = \frac{n}{(1-p)^n} \int_p^1 \Big( (s-p)^n + p(s-p)^{n-1} \Big) \,\dd s
    = \frac{n}{(1-p)^n}\left( \frac{(1-p)^{n+1}}{n+1} + \frac{p}{n}(1-p)^n \right)\\
  & = \frac{p}{n+1} + \frac{n}{n+1}, \qquad p\in[0,1).
 \end{align*}
So for each \ $\vare\in(0,1)$, \ the inequality \ $\ES_{X,n}(0)\leq  \VaR_X(1-\vare)$ \ is equivalent
 to \ $\frac{n}{n+1}\leq 1-\vare$, \ and hence, by Proposition \ref{Pro1}, if \ $\vare\in(0,\frac{1}{n+1}]$, \ then
 \ $\Pi_{\vare,n}(X)$ \ is a solution of the equation \ $\ES_{X,n}(1-c\vare) = \VaR_X(1-\vare)$, \ $c\in[1,\frac{1}{\vare}]$, \ taking the form
 \[
   \frac{1-c\vare}{n+1} +\frac{n}{n+1} = 1-\vare,\quad c\in\left[1,\frac{1}{\vare}\right].
 \]
Hence \ $\Pi_{\vare,n}(X) = n+1$ \ for \ $\vare\in(0,\frac{1}{n+1}]$.
\ If \ $\vare\in(\frac{1}{n+1},1)$, \ then, by Definition \ref{Def_PELVE2}, \ $\Pi_{\vare,n}(X) = \infty$.
\ Note that, by Li and Wang \cite[part (i) of Example 5]{LiWan}, \ $\Pi_{\vare}(X)=2$ \ for \ $\vare\in(0,\frac{1}{2}]$, \ and hence
 \ $\Pi_{\vare,n}(X) > \Pi_{\vare}(X)$ \ for \ $\vare\in(0,\frac{1}{n+1}]$ \ (as it is expected, see part (ii) of Remark \ref{Rem2}).

Let \ $Y$ \ be a random variable with uniform distribution on the interval \ $[a,b]$, \ where \ $a<b$, \ $a,b\in\RR$.
\ Using that the distribution of \ $Y$ \ coincides with that of \ $(b-a)X+a$, \ part (i) of Theorem \ref{Thm_PELVE_tulajdonsagok} yields
 \ $\Pi_{\vare,n}(Y) = \Pi_{\vare,n}((b-a)X+a) = \Pi_{\vare,n}(X)$, \ so
 \[
   \Pi_{\vare,n}(Y)
     = \begin{cases}
         {n+1} & \text{if \ $\vare\in(0,\frac{1}{n+1}]$,}\\
         \infty & \text{if \ $\vare\in(\frac{1}{n+1},1)$.}
       \end{cases}
 \]
 \proofend
\end{Ex}

\begin{Ex}[Exponential distribution]\label{Ex_Exponencialis}
Let \ $X$ \ be an exponentially distributed random variable with parameter \ $1$, \ and let \ $n\in\NN$.
\ Then \ $\VaR_X(p) = -\ln(1-p)$, \ $p\in(0,1)$, \ and
\begin{align*}
 \ES_{X,n}(p)
  & = \frac{n}{(1-p)^n} \int_p^1 (s-p)^{n-1}\VaR_X(s)\,\dd s
    = -\frac{n}{(1-p)^n} \int_p^1 (s-p)^{n-1} \ln(1-s)\,\dd s\\
  & = -\frac{n}{(1-p)^n} \int_0^{1-p} (1-p-s)^{n-1} \ln(s)\,\dd s,
  \qquad p\in [0,1).
 \end{align*}
Using formula 2.725/2 in Gradshteyn and Ryzhik \cite{GraRyz} with \ $a:=1-p$, \ $b:=-1$ \ and \ $m:=n-1$, \ we have
 \begin{align*}
  \int (1-p-s)^{n-1} \ln(s)\,\dd s
    & = -\frac{1}{n}\Big( (1-p-s)^n - (1-p)^n \Big)\ln(s) \\
    &\phantom{=\;} - \sum_{k=0}^{n-1} \binom{n-1}{k}\frac{(-1)^k}{(k+1)^2} (1-p)^{n-1-k}s^{k+1}
              + C,
 \end{align*}
 where \ $C\in\RR$.
\ It yields that
 \begin{align*}
  \ES_{X,n}(p)
    = -\ln(1-p) + n \sum_{k=0}^{n-1} \binom{n-1}{k}\frac{(-1)^k}{(k+1)^2}
    = H_n - \ln(1-p), \qquad p\in[0,1),
 \end{align*}
 where \ $H_n:=\sum_{k=1}^n \frac{1}{k}$ \ denotes the $n$-th harmonic number.
Indeed, the second equality above follows by
 \begin{align*}
  n \sum_{k=0}^{n-1} \binom{n-1}{k}\frac{(-1)^k}{(k+1)^2}
  & = \sum_{k=0}^{n-1} \frac{n!}{k!(n-1-k)!}\frac{(-1)^k}{(k+1)^2}
    = \sum_{k=0}^{n-1} \binom{n}{k}(n-k)\frac{(-1)^k}{(k+1)^2} \\
  & = \sum_{k=1}^n \binom{n}{k-1}(n-k+1)\frac{(-1)^{k-1}}{k^2}
   = \sum_{k=1}^n \frac{n!}{(k-1)!(n-k)!}\frac{(-1)^{k-1}}{k^2}\\
  & = - \sum_{k=1}^n \binom{n}{k}\frac{(-1)^k}{k}
   = H_n,
 \end{align*}
 where the last equality is a well-known identity for \ $H_n$.
\ In particular, we have
 \begin{align*}
  \ES_{X,2}(p) = - \ln(1-p) + \frac{3}{2}, \qquad p\in[0,1).
 \end{align*}

So for each \ $\vare\in(0,1)$, \ the inequality \ $\ES_{X,n}(0) \leq  \VaR_X(1-\vare)$ \ is equivalent to
 \ $H_n \leq -\ln(\vare)$, \ i.e., \ $\vare\in(0, \ee^{-H_n}]$.
\ Hence, by Proposition \ref{Pro1}, if \ $\vare\in(0, \ee^{-H_n}]$,
 \ then \ $\Pi_{\vare,n}(X)$ \ is a solution of the equation \ $\ES_{X,n}(1-c\vare) = \VaR_X(1-\vare)$, \ $c\in[1,\frac{1}{\vare}]$, \ taking the form
 \[
   -\ln(c\vare)  + H_n = -\ln(\vare),\quad c\in\left[1,\frac{1}{\vare}\right].
 \]
Hence
 \[
  \Pi_{\vare,n}(X) = \ee^{H_n} \qquad \text{for \ $\vare\in(0, \ee^{-H_n}]$.}
 \]
If \ $\vare\in(\ee^{-H_n},1)$, \ then, by Definition \ref{Def_PELVE2}, \ $\Pi_{\vare,n}(X) = \infty$

In particular, we have \ $\Pi_{\vare,2}(X) = \ee^{\frac{3}{2}}\approx 4.482$ \ for \ $\vare\in(0,\ee^{-3/2}]$,
 \ and \ $\Pi_{\vare,2}(X) = \infty$ \ for \ $\vare\in(\ee^{-3/2},1)$.

Note that, by Li and Wang \cite[Example 5, part (ii)]{LiWan}, \ $\Pi_{\vare}(X) = \ee$ \ for \ $\vare\in(0,\ee^{-1}]$, \
  so \ $\Pi_{\vare,n}(X) > \Pi_{\vare}(X)$ \ for \ $\vare\in(0, \ee^{-H_n} ]$ \
  (as it is expected, see part (ii) of Remark \ref{Rem2}).

Let \ $Y$ \ be an exponentially distributed random variable with parameter \ $\lambda>0$.
\ Using that the distribution of \ $Y$ \ coincides with that of \ $\frac{1}{\lambda} X$, \ part (i) of Theorem \ref{Thm_PELVE_tulajdonsagok} yields
 \ $\Pi_{\vare,n}(Y) = \Pi_{\vare,n}(\frac{1}{\lambda}X) = \Pi_{\vare,n}(X)$, \ so
\[
   \Pi_{\vare,n}(Y)
     = \begin{cases}
         \ee^{H_n}
                   & \text{if \ $\vare\in\big(0, \ee^{-H_n} \big]$,}\\[1mm]
         \infty & \text{if \ $\vare\in\big(\ee^{-H_n},1\big)$.}
       \end{cases}
 \]
 \proofend
\end{Ex}

\begin{Ex}[Normal distribution]
Let \ $X$ \ be a standard normally distributed random variable.
Let \ $\Phi$ \ and \ $\varphi$ \ denote the distribution function and density function of \ $X$, \ respectively.
\ Then \ $\VaR_X(p) = \Phi^{-1}(p)$, \ $p\in(0,1)$, \ and, by substitution \ $s=\Phi(x)$, \ we have
 \begin{align*}
 \ES_{X,2}(p)
  & = \frac{2}{(1-p)^2} \int_p^1 (s-p)\VaR_X(s)\,\dd s
    = \frac{2}{(1-p)^2} \int_p^1 (s-p) \Phi^{-1}(s)\,\dd s\\
  & =\frac{2}{(1-p)^2} \int_{\Phi^{-1}(p)}^\infty (\Phi(x)-p) x\varphi(x)\,\dd x\\
  & = \frac{2}{(1-p)^2} \left( \int_{\Phi^{-1}(p)}^\infty  x\Phi(x)\varphi(x) \,\dd x  - p \int_{\Phi^{-1}(p)}^\infty  x\varphi(x) \,\dd x  \right),
      \qquad p\in [0,1),
 \end{align*}
 where \ $\Phi^{-1}(0)$ \ is defined to be \ $-\infty$.
\ Here, since \ $\varphi(x)=\frac{1}{\sqrt{2\pi}}\ee^{-\frac{x^2}{2}}$, \ $x\in\RR$, \ we have
 \[
  - p \int_{\Phi^{-1}(p)}^\infty  x\varphi(x) \,\dd x
    = p [\varphi(x)]_{\Phi^{-1}(p)}^\infty
    = -p \varphi(\Phi^{-1}(p)),\qquad p\in [0,1).
 \]
Further, by partial integration and then substitution \ $x=\frac{y}{\sqrt{2}}$, \ we get
 \begin{align*}
   \int_{\Phi^{-1}(p)}^\infty  x\Phi(x)\varphi(x) \,\dd x
     &= [-\Phi(x)\varphi(x)]_{\Phi^{-1}(p)}^\infty + \int_{\Phi^{-1}(p)}^\infty \varphi^2(x)\,\dd x\\
     &= p \varphi(\Phi^{-1}(p)) + \int_{\Phi^{-1}(p)}^\infty \frac{1}{2\pi} \ee^{-x^2} \,\dd x\\
     &= p \varphi(\Phi^{-1}(p)) + \frac{1}{2\sqrt{\pi}} \int_{\sqrt{2}\Phi^{-1}(p)}^\infty \frac{1}{\sqrt{2\pi}} \ee^{-\frac{y^2}{2}} \,\dd y\\
     &= p \varphi(\Phi^{-1}(p)) + \frac{1}{2\sqrt{\pi}}(1-\Phi\big(\sqrt{2} \Phi^{-1}(p)\big)),
     \qquad p\in[0,1).
 \end{align*}
Hence
 \[
    \ES_{X,2}(p) =  \frac{1}{\sqrt{\pi}(1-p)^2}(1-\Phi\big(\sqrt{2} \Phi^{-1}(p)\big)), \qquad p\in[0,1).
 \]
So for each \ $\vare\in(0,1)$, \ the inequality \ $\ES_{X,2}(0) \leq  \VaR_X(1-\vare)$ \ is equivalent
 to \ $\frac{1}{\sqrt{\pi}}\leq \Phi^{-1}(1-\vare)$, \ i.e., \ $\vare\in(0, 1 - \Phi(1/\sqrt{\pi})]$, \
  and hence, by Proposition \ref{Pro1}, if \ $\vare\in(0, 1 - \Phi(1/\sqrt{\pi})]\approx (0,0.286]$, \ then
 \ $\Pi_{\vare,2}(X)$ \ is a solution of the equation \ $\ES_{X,2}(1-c\vare) = \VaR_X(1-\vare)$, \ $c\in[1,\frac{1}{\vare}]$, \ taking the form
 \begin{align}\label{help_ES_2_eq_norm}
     \frac{1}{\sqrt{\pi}c^2\vare^2}(1-\Phi\big(\sqrt{2} \Phi^{-1}(1-c\vare)\big))
        = \Phi^{-1}(1-\vare),\quad c\in\left[1,\frac{1}{\vare}\right].
 \end{align}
If \ $\vare\in(1-\Phi(1/\sqrt{\pi}),1)$, \ then, by Definition \ref{Def_PELVE2}, \ $\Pi_{\vare,2}(X) = \infty$.

Let \ $Y$ \ be a normally distributed random variable with mean \ $m$ \ and variance \ $\sigma^2$, \ where \ $m\in\RR$ \ and \ $\sigma>0$.
Using that the distribution of \ $Y$ \ coincides with that of \ $\sigma X + m$, \ part (i) of Theorem \ref{Thm_PELVE_tulajdonsagok} yields
 \ $\Pi_{\vare,2}(Y) = \Pi_{\vare,2}(\sigma X+ m) = \Pi_{\vare,2}(X)$.

Using the software R, for levels \ $\vare\in\{ 0.1,\, 0.05, \, 0.01,\, 0.005\}$ \ we calculate an approximated value of
 the unique root \ $\Pi_{\vare,2}(X)$ \ of the equation \eqref{help_ES_2_eq_norm}, see Table \ref{Table1}.
   \begin{table}[ht]
   \begin{center}
		\begin{tabular}{ |c|c| }
			\hline
			$\vare$ & $\PELVES$ \ of \ $\cN(m,\sigma^2)$ \\
			\hline
			0.100 & 3.92217 \\
			0.050 & 4.04082 \\
			0.010 & 4.18527 \\
			0.005 & 4.22188 \\
			\hline
		\end{tabular}
      \caption{Approximations of \ $\PELVES$-values of normal distribution  $\cN(m,\sigma^2)$ \ ($m\in\RR$, \ $\sigma>0$).}
      \label{Table1}
	\end{center}
  \end{table}
 \proofend
\end{Ex}

\begin{Ex}[Pareto distribution]\label{Ex_Pareto}
Let \ $X$ \ be a random variable with Pareto distribution having parameters \ $k>0$ \ and \ $\alpha>0$, \ i.e.,
 the distribution function of \ $X$ \ takes the form \ $F_X:\RR\to[0,1]$,
 \[
   F_X(x):=\begin{cases}
             1 - \left(\frac{k}{x}\right)^\alpha  & \text{if \ $x\geq k$,}\\
             0  & \text{if \ $x<k$.}
           \end{cases}
 \]
Further, let \ $n\in\NN$.
\ Then
  \[
     \VaR_X(p) = k(1-p)^{-\frac{1}{\alpha}},\qquad p\in(0,1).
  \]
In what follows we suppose that \ $\alpha>1$, \ yielding that \ $X\in L^1$.
\ In this case we have
\begin{align*}
 \ES_{X,n}(p)
  & = \frac{n}{(1-p)^n} \int_p^1 (s-p)^{n-1}\VaR_X(s)\,\dd s
    = \frac{n}{(1-p)^n} \int_p^1 (s-p)^{n-1} k (1-s)^{-\frac{1}{\alpha}} \,\dd s\\
  & = \frac{kn}{(1-p)^n} \int_0^{1-p} (1-p-s)^{n-1} s^{-\frac{1}{\alpha}} \,\dd s
\end{align*}
\begin{align*}
  & = \frac{kn}{(1-p)^n} \int_0^{1-p} \sum_{j=0}^{n-1} \binom{n-1}{j} (-s)^j (1-p)^{n-1-j} s^{-\frac{1}{\alpha}} \,\dd s\\
  & = \frac{kn}{(1-p)^n} \sum_{j=0}^{n-1} \binom{n-1}{j} (1-p)^{n-1-j} (-1)^j \int_0^{1-p} s^{j-\frac{1}{\alpha}} \,\dd s\\
  & = \frac{kn}{(1-p)^n} \sum_{j=0}^{n-1} \binom{n-1}{j} (1-p)^{n-1-j} (-1)^j \frac{(1-p)^{j-\frac{1}{\alpha}+1}}{j-\frac{1}{\alpha}+1}\\
  & = kn (1-p)^{-\frac{1}{\alpha}} \sum_{j=0}^{n-1} \binom{n-1}{j} \frac{(-1)^j }{j-\frac{1}{\alpha}+1},
   \qquad p\in[0,1).
 \end{align*}
In particular, we have
 \begin{align*}
 \ES_{X,2}(p) = \frac{2k\alpha^2}{(\alpha-1)(2\alpha-1)} (1-p)^{-\frac{1}{\alpha}},
   \qquad p\in[0,1).
 \end{align*}
So for each \ $\vare\in(0,1)$, \ the inequality \ $\ES_{X,n}(0) \leq  \VaR_X(1-\vare)$ \ is equivalent
 to
 \[
 kn \sum_{j=0}^{n-1} \binom{n-1}{j} \frac{(-1)^j }{j-\frac{1}{\alpha}+1}
       \leq k \vare^{-\frac{1}{\alpha}}, \qquad \text{i.e.,}\qquad
   \vare\in\left(0, \left( n \sum_{j=0}^{n-1} \binom{n-1}{j} \frac{(-1)^j }{j-\frac{1}{\alpha}+1} \right)^{-\alpha} \right],
  \]
  and hence, by Proposition \ref{Pro1}, if \ $\vare\in\Big(0, \left( n \sum_{j=0}^{n-1} \binom{n-1}{j} \frac{(-1)^j }{j-\frac{1}{\alpha}+1} \right)^{-\alpha} \Big]$, \ then
 \ $\Pi_{\vare,n}(X)$ \ is a solution of the equation \ $\ES_{X,n}(1-c\vare) = \VaR_X(1-\vare)$, \ $c\in[1,\frac{1}{\vare}]$, \ taking the form
 \[
   kn (c\vare)^{-\frac{1}{\alpha}} \sum_{j=0}^{n-1} \binom{n-1}{j} \frac{(-1)^j }{j-\frac{1}{\alpha}+1} = k \vare^{-\frac{1}{\alpha}},
   \qquad c\in\Big[1,\frac{1}{\vare}\Big].
 \]
Hence
\begin{align*}
    \Pi_{\vare,n}(X) = \left( n \sum_{j=0}^{n-1} \binom{n-1}{j} \frac{(-1)^j }{j-\frac{1}{\alpha}+1}  \right)^\alpha
    \qquad \text{\ for \ $\vare\in\left(0,\left( n \sum_{j=0}^{n-1} \binom{n-1}{j} \frac{(-1)^j }{j-\frac{1}{\alpha}+1} \right)^{-\alpha} \right]$.}
 \end{align*}
If \ $\vare\in\Big(\left( n \sum_{j=0}^{n-1} \binom{n-1}{j} \frac{(-1)^j }{j-\frac{1}{\alpha}+1} \right)^{-\alpha},1\Big)$, \ then, by Definition \ref{Def_PELVE2}, \ $\Pi_{\vare,n}(X) = \infty$.

In particular, we have
 \begin{align}\label{help_PELVE_7}
    \Pi_{\vare,2}(X) = \left(\frac{2\alpha^2}{(\alpha-1)(2\alpha-1)}\right)^\alpha
    \qquad \text{\ for \ $\vare\in\left(0,\left(\frac{(\alpha-1)(2\alpha-1)}{2\alpha^2}\right)^\alpha\right]$,}
 \end{align}
 and if \ $\vare\in\Big(\Big(\frac{(\alpha-1)(2\alpha-1)}{2\alpha^2}\Big)^\alpha,1\Big)$, \ then \ $\Pi_{\vare,2}(X) = \infty$.
\ Note that
 \[
    \lim_{x\to\infty} \left(\frac{(x-1)(2x-1)}{2x^2}\right)^x
      = \lim_{x\to\infty}  \left(1-\frac{1}{x}\right)^x \left[ \left(1 - \frac{1}{2x}\right)^{2x} \right]^{\frac{1}{2}}
      = \ee^{-1}\ee^{-\frac{1}{2}} = \ee^{-\frac{3}{2}},
 \]
 and the function
 \ $(1,\infty)\ni x\mapsto \left(\frac{2x^2}{(x-1)(2x-1)}\right)^x$ \ is monotone decreasing.
Indeed,
 \[
   \left(\frac{2x^2}{(x-1)(2x-1)}\right)^x
     = \left(\frac{x}{x-1}\right)^x
        \left(\left(\frac{2x}{2x-1}\right)^{2x}\right)^{\frac{1}{2}},
        \qquad x\in(1,\infty),
 \]
 and the logarithm of the function $(1,\infty)\ni x\mapsto (x/(x-1))^x$, \ i.e.,
 the function \ $(1,\infty)\ni x\mapsto x \ln(x/(x-1))$, \ is monotone decreasing.
Hence, by \eqref{help_PELVE_7}, for each \ $\vare\in\left(0,\left(\frac{(\alpha-1)(2\alpha-1)}{2\alpha^2}\right)^\alpha \right]$, \ we have
 \begin{align}\label{help_PELVE_8}
 \Pi_{\vare,2}(X) \geq
      \lim_{x\to\infty} \left(\frac{2x^2}{(x-1)(2x-1)}\right)^x = \ee^{\frac{3}{2}}\approx 4,482,
 \end{align}
 where, by Example \ref{Ex_Exponencialis},  the limit \ $\ee^{\frac{3}{2}}$ \ is nothing else but the \ $\PELVES$-value (at the given level \ $\vare$) \
 of an exponentially distributed random variable.
The inequality \eqref{help_PELVE_8} for \ $\PELVES$ \ can be considered as the counterpart of the corresponding inequality for \ $\PELVE$ \ of \ $X$ \
 due to Li and Wang \cite[inequality (8)]{LiWan}.
For the parameters \ $\alpha\in\{ 2,\, 10,\, 30\}$, \ we calculate \ $\Pi_{\vare,2}(X)$, \ see Table \ref{Table2}.
 \begin{table}[ht]
   \begin{center}
 \begin{tabular}{ |c|c|c|c|c| }
    \hline
   $\PELVES$  & $\vare\in\big(0, \frac{9}{64}\big]$  &  $\vare\in\Big(\frac{9}{64}, \big(\frac{171}{200}\big)^{10} \Big]$
                  & $\vare\in\Big(\big(\frac{171}{200}\big)^{10} , \big(\frac{1711}{1800}\big)^{30} \Big]$
                  & $\vare\in\Big( \big(\frac{1711}{1800}\big)^{30} , 1\Big)$ \\
  \hline
  Pareto(k,2) & 7.112 & $\infty$ & $\infty$ & $\infty$ \\
  \hline
  Pareto(k,10) & 4.791 & 4.791 & $\infty$ & $\infty$ \\
  \hline
  Pareto(k,30) & 4.578 & 4.578 & 4.578 & $\infty$ \\
 \hline
 \end{tabular}
 \caption{$\PELVES$-values of Pareto distribution Pareto$(k,\alpha)$ \ ($k>0$, \ $\alpha\in\{2,10,30\}$), rounded up to 3 decimal places.}
 \label{Table2}
 \end{center}
  \end{table}
\proofend
\end{Ex}

\section{Generalized Pareto distributions and \ $\PELVES$}\label{Sec_PELVE2_Gen_Par}

In this section, we calculate the $\PELVE_2$-values of random variables with some generalized Pareto distribution and
 random variables of which the excess distribution function over a threshold is given by the distribution
 function of a generalized Pareto distribution with tail index less than \ $1$.
\ It will turn out that the $\PELVE_2$-value at a level \ $\epsilon$ \ of such random variables
 depends on the tail index but not on \ $\vare$ \ (below some threshold).
Such a result was already established for the \ $\PELVE$-values and conditional $\PELVE$-values
 of the random variables in question by Fiori and Rosazza Gianin \cite[Proposition 15]{FioGia}.
Both results might be useful for estimating the tail index of the random variables in question.

First, we recall the notion of a generalized Pareto distribution.

\begin{Def}\label{Def_Gen_Par}
Let \ $\kappa\in\RR$ \ and \ $\beta>0$.
\ We say that a random variable \ $X$ \ has a generalized Pareto distribution with parameters \ $\kappa$ \ and \ $\beta$
 \ if its distribution function \ $F_X$ \ takes the form:
 \begin{itemize}
   \item in case of \ $\kappa>0$,
           \[
               F_X(x):=\begin{cases}
                              1 - \left(1+\frac{\kappa}{\beta}x\right)^{-\frac{1}{\kappa}}  & \text{if \ $x\geq 0$,}\\
                              0 & \text{if \ $x<0$,}
                             \end{cases}
           \]
   \item in case of \ $\kappa=0$,
         \[
               F_X(x):=\begin{cases}
                              1 - \ee^{-\frac{x}{\beta}}  & \text{if \ $x\geq 0$,}\\
                              0 & \text{if \ $x<0$,}
                             \end{cases}
           \]
    \item in case of \ $\kappa<0$,
          \[
               F_X(x):=\begin{cases}
                              1 & \text{if \ $x>-\frac{\beta}{\kappa}$,}\\
                              1 - \left(1+\frac{\kappa}{\beta}x\right)^{-\frac{1}{\kappa}}  & \text{if \ $x\in[0,-\frac{\beta}{\kappa}]$,}\\
                              0 & \text{if \ $x<0$.}
                             \end{cases}
           \]
 \end{itemize}
The parameter \ $\kappa$ \ is sometimes called the tail index of \ $X$.
\ The distribution function of a random variable having generalized Pareto distribution
 with parameters \ $\kappa\in\RR$ \ and \ $\beta>0$ \ is denoted by \ $G_{\kappa,\beta}$.
\end{Def}

\begin{Rem}
If \ $X$ \ is a random variable having a generalized Pareto distribution with parameters \ $\kappa>0$ \ and \ $\beta>0$,
 \ then \ $X+\frac{\beta}{\kappa}$ \ has a (usual) Pareto distribution with parameters \ $\frac{\beta}{\kappa}$ \ and \ $\frac{1}{\kappa}$
 \ (recalled in Example \ref{Ex_Pareto}).
\ Further, if \ $X$ \ is a random variable having a generalized Pareto distribution with parameters \ $\kappa=0$ \ and \ $\beta>0$,
 \ then \ $X$ \ is in fact exponentially distributed with parameter \ $\frac{1}{\beta}$.
   \proofend
\end{Rem}

Next, to give an application of Theorem \ref{Thm_PELVE_tulajdonsagok}, we calculate the $\PELVES$-value
 of a generalized Pareto distribution with parameters \ $\kappa\in(0,1)$ \ and \ $\beta:=1$.

\begin{Ex}
Let \ $\kappa\in(0,1)$ \ and let \ $X$ \ be a random variable with distribution function \ $F_X:\RR\to[0,1]$,
 \[
   F_X(x):=\begin{cases}
             1 - (1+\kappa x)^{-\frac{1}{\kappa}}  & \text{if \ $x\geq 0$,}\\
             0  & \text{if \ $x<0$.}
           \end{cases}
 \]
Then \ $X$ \ has a generalized Pareto distribution with parameters \ $\kappa$ \ and \ $\beta:=1$,
 \ and \ $X$ \ has the unbounded support \ $[0,\infty)$.
\ Consequently, \ $X+\frac{1}{\kappa}$ \ has a (usual) Pareto distribution with parameters \ $\frac{1}{\kappa}$ \
  and \ $\frac{1}{\kappa}$.
This yields that the calculations of \ $\VaR$, \ $2^{\mathrm{nd}}$-order \ $\ES$ \ and \ $\PELVE_2$ \ of \ $X$ \
 can be traced back to those of \ $X+\frac{1}{\kappa}$, \ for which we can use Example \ref{Ex_Pareto}.

Using Example \ref{Ex_Pareto} and the translation invariance of \ $\VaR$, \ we have
  \[
     \VaR_X(p) = \VaR_{X+\frac{1}{\kappa}}(p) - \frac{1}{\kappa}
               = \frac{1}{\kappa}(1-p)^{-\kappa} - \frac{1}{\kappa}
               = \frac{-1+(1-p)^{-\kappa}}{\kappa}  ,\qquad p\in(0,1).
  \]
Since \ $\kappa\in(0,1)$, \ we have \ $X+\frac{1}{\kappa}\in L^1$, \ and hence \ $X\in L^1$.
\ Consequently, using again Example \ref{Ex_Pareto} and the translation invariance of the \ $2^{\mathrm{nd}}$-order
 \ $\ES$ \ (being a coherent risk measure on \ $L^1$, \ see part (ii) of Remark \ref{Rem4}), we have
 \begin{align*}
 \ES_{X,2}(p)
  & = \ES_{X+\frac{1}{\kappa},2}(p) - \frac{1}{\kappa}
    = \frac{2}{\kappa^3\left(\frac{1}{\kappa} - 1 \right)\left(\frac{2}{\kappa} - 1 \right)}(1-p)^{-\kappa} -  \frac{1}{\kappa} \\
  & = -\frac{1}{\kappa} + \frac{2(1-p)^{-\kappa}}{\kappa(1-\kappa)(2-\kappa)},
     \qquad p\in[0,1).
 \end{align*}
Consequently, for each \ $\vare\in(0,1)$, \ the inequality \ $\ES_{X,2}(0) \leq  \VaR_X(1-\vare)$ \ is
 equivalent to
 \[
     -\frac{1}{\kappa} + \frac{2}{\kappa(1-\kappa)(2-\kappa)}
        \leq \frac{-1+\vare^{-\kappa}}{\kappa}, \qquad \text{i.e.,}\qquad
      \vare\in\left(0, \left( \frac{(1-\kappa)(2-\kappa)}{2} \right)^{\frac{1}{\kappa}} \right].
  \]
Using part (i) of Theorem \ref{Thm_PELVE_tulajdonsagok}
 and \eqref{help_PELVE_7} yield that
 \[
    \Pi_{\vare,2}(X)
     = \Pi_{\vare,2}\Big(X+\frac{1}{\kappa}\Big) = \left(\frac{2}{(1-\kappa)(2-\kappa)}\right)^{\frac{1}{\kappa}}
    \qquad \text{\ for \ $\vare\in\left(0,\left( \frac{(1-\kappa)(2-\kappa)}{2} \right)^{\frac{1}{\kappa}} \right]$.}
 \]
If \ $\vare\in\Big(\Big( \frac{(1-\kappa)(2-\kappa)}{2} \Big)^{\frac{1}{\kappa}} ,1\Big)$, \ then, by Definition \ref{Def_PELVE2}, \ $\Pi_{\vare,2}(X) = \infty$.

Finally, note that, by L'Hospital's rule, we get
 \begin{align*}
    \lim_{\kappa\downarrow 0} \left( -\frac{1}{\kappa} + \frac{2(1-p)^{-\kappa}}{\kappa(1-\kappa)(2-\kappa)} \right)
       & = \lim_{\kappa\downarrow 0} \frac{-(1-\kappa){ (2-\kappa)}+2(1-p)^{-\kappa}}{\kappa(1-\kappa)(2-\kappa)}\\
       & = \frac{1}{2}\lim_{\kappa\downarrow 0}\frac{-\kappa^2 + 3\kappa -2 + 2(1-p)^{-\kappa}}{\kappa} \\
       & = \frac{1}{2}\lim_{\kappa\downarrow 0} \Big(-2\kappa + 3 - 2(1-p)^{-\kappa}\ln(1-p) \Big)\\
       & = -\ln(1-p) + \frac{3}{2},\qquad p\in[0,1),
 \end{align*}
 where \ $-\ln(1-p) + \frac{3}{2}$ \ is nothing else but the \ $2^{\mathrm{nd}}$-order Expected Shortfall of an exponentially distributed random variable at a level \ $p$,
 \  see Example \ref{Ex_Exponencialis}.
This is in accordance with in accordance with Lemma \ref{Lem_ES2_conv}, since
 a generalized Pareto distribution having parameters \ $\kappa>0$ \ and \ $1$ \ converges in distribution to the
 exponential distribution with parameter $1$ as $\kappa\downarrow 0$.
\proofend
\end{Ex}

For a random variable \ $X$ \  with distribution function \ $F_X$,
 let \ $x_{F_X}$ \ denote the right endpoint of \ $F_X$, \ i.e., \ $x_{F_X}:=\sup\{x\in\RR : F_X(x)<1\}$.
\ If \ $X$ \ has a generalized Pareto distribution with parameters \ $\kappa\in\RR$ \ and \ $\beta>0$, \ then
 \[
   x_{F_X}=\begin{cases}
           \infty   & \text{if \ $\kappa\geq 0$,}\\
           -\frac{\beta}{\kappa}  & \text{if \ $\kappa< 0$.}
        \end{cases}
 \]

\begin{Def}
Let \ $X$ \ be a non-negative random variable with distribution function \ $F_X$.
\ The excess distribution function corresponding to \ $F_X$ \ over a threshold \ $u\in[0,x_{F_X})$ \ is given by \ $F_{X,u}:[0,\infty)\to\RR$,
 \[
   F_{X,u}(x):=\PP(X-u\leq x \mid X>u)
          = \frac{F_X(x+u) - F_X(u)}{1-F_X(u)},
          \qquad x\geq 0.
 \]
Note that if \ $x_{F_X}<\infty$, \ then \ $F_{X,u}(x)=1$ \ for \ $x\geq x_{F_X}-u$.
\end{Def}

For the forthcoming Propositions \ref{Pro_GPD_1}, \ref{Pro_GPD_2} and \ref{Pro_GPD_3},
 one can refer to Example 5.19, Lemma 5.22,
 formulae (5.18), (5.19) and (5.20) in McNeil et al.\ \cite{McnFreEmb}.

\begin{Pro}\label{Pro_GPD_1}
Let \ $X$ \ be a random variable having a generalized Pareto distribution with parameters \ $\kappa\in\RR$ \ and \ $\beta>0$.
\ For the excess distribution function \ $F_{X,u}$ \ corresponding to \ $F_X$ \ over a threshold \ $u$, \ we have
 \ $F_{X,u}(x) = G_{\kappa,\beta+\kappa u}(x)$ \ for \ $x\geq 0$ \ and \ $u\geq 0$ \ in case of \ $\kappa\geq 0$;
 \ and for \ $x\geq 0$ \ and \ $u<-\frac{\beta}{\kappa}$ \ in case of \ $\kappa<0$.
\end{Pro}

\begin{Pro}\label{Pro_GPD_2}
Let \ $X$ \ be a non-negative random variable, and assume that there exist \ $u\in[0,x_{F_X})$, \ $\kappa\in\RR$ \ and \ $\beta>0$ \ such that
 \ $F_{X,u}(x) = G_{\kappa,\beta}(x)$ \ for \ $x\in[0,x_{F_X}-u)$.
Then \ $F_{X,v}(x) = G_{\kappa,\beta+\kappa(v-u)}(x)$ \ for \ $x\geq 0$ \ and \ $v\geq u$ \ in case of \ $\kappa\geq 0$; \
 and for \ $x\in[0,-\frac{\beta}{\kappa}-(v-u))$ \ and \ $v\in[u,u-\frac{\beta}{\kappa})$ \ in case of \ $\kappa<0$.
\end{Pro}

\begin{Pro}\label{Pro_GPD_3}
Let \ $X$ \ be a non-negative random variable, and assume that there exist \ $u\in[0,x_{F_X})$, \ $\kappa\in\RR$ \ and \ $\beta>0$ \ such that
 \ $F_{X,u}(x) = G_{\kappa,\beta}(x)$ \ for \ $x\in[0,x_{F_X}-u)$.
Then for each \ $p\in(F_X(u),1)$, \ we have
 \begin{align}\label{help_VaR_GDP}
  \VaR_X(p)=\begin{cases}
          u + \frac{\beta}{\kappa}\left( \left( \frac{1-p}{1-F_X(u)}\right)^{-\kappa} - 1 \right)  & \text{in case of \ $\kappa\ne0$,}\\
          u - \beta\ln\left(\frac{1-p}{1-F_X(u)}\right) & \text{in case of \ $\kappa=0$.}
            \end{cases}
 \end{align}
If, in addition \ $\kappa<1$, \ then we have
 \[
   \ES_X(p) = \frac{1}{1-\kappa}\VaR_X(p) + \frac{\beta - \kappa u}{1-\kappa},\qquad p\in(F_X(u),1),
 \]
 and
 \begin{align}\label{help_ES_VAR_ratio}
   \lim_{p\uparrow 1} \frac{\ES_X(p)}{\VaR_X(p)}
      = \begin{cases}
          \frac{1}{1-\kappa} & \text{in case of \ $\kappa\in[0,1)$,}\\
          1 & \text{in case of \ $\kappa<0$.}
        \end{cases}
 \end{align}
\end{Pro}

In the next proposition, we calculate the $\PELVE_2$ of a non-negative random variable having a generalized Pareto
 excess distribution function.
This result can be considered as a counterpart of the corresponding results for $\PELVE$ and conditional $\PELVE$
 (see Definition \ref{Def_Fiori_Gianin}) in Proposition 15 in Fiori and Rosazza Gianin \cite{FioGia}.

\begin{Pro}
Let \ $X$ \ be a non-negative random variable, and assume that there exist \ $u\in[0,x_{F_X})$, \ $\kappa<1$ \ and \ $\beta>0$ \ such that
 \ $F_{X,u}(x) = G_{\kappa,\beta}(x)$ \ for \ $x\in[0,x_{F_X}-u)$.
\ In case of \ $\kappa<1$ \ with \ $\kappa\ne0$, \ we have
 \[
   \Pi_{\vare,2}(X) = \left(\frac{2}{(1-\kappa)(2-\kappa)}\right)^{\frac{1}{\kappa}}
       \qquad \text{for \ $0<\vare<(1-F_X(u))\left(\frac{(1-\kappa)(2-\kappa)}{2}\right)^{\frac{1}{\kappa}}$,}
 \]
 and in case of \ $\kappa=0$, \ we have
 \[
   \Pi_{\vare,2}(X) = \ee^{\frac{3}{2}}
       \qquad \text{for \ $0<\vare<(1-F_X(u))\ee^{-\frac{3}{2}}$.}
 \]
Further,
 \begin{align}\label{help_ES2_VAR_ratio}
  \lim_{p\uparrow 1} \frac{\ES_{X,2}(p)}{\VaR_X(p)}
       = \begin{cases}
            \frac{2}{(1-\kappa)(2-\kappa)} & \text{if \ $\kappa\in[0,1)$,}\\
            1 & \text{if \ $\kappa<0$.}
         \end{cases}
 \end{align}
\end{Pro}

\noindent{\bf Proof.}
Recall that \ $\VaR_X(p)$, \ $p\in(F_X(u),1)$, \ is given in \eqref{help_VaR_GDP}.
Next, we calculate \ $\ES_{X,2}(p)$ \ for \ $p\in(F_X(u),1)$.

First, we consider the case of \ $\kappa=0$. \
For \ $p\in(F_X(u),1)$ \ we have
 \begin{align*}
   \ES_{X,2}(p)
    & = \frac{2}{(1-p)^2} \int_p^1(s-p)\left(u - \beta\ln\left(\frac{1-s}{1-F_X(u)}\right)\right)\dd s   \\
    & = \frac{2}{(1-p)^2}\Big( u+ \beta\ln(1-F_X(u))\Big) \int_p^1 (s-p)\,\dd s
       - \frac{2\beta}{(1-p)^2}\int_p^1 s\ln(1-s)\,\dd s\\
    &\phantom{=\;}   +\frac{2\beta p}{(1-p)^2}\int_p^1\ln(1-s)\,\dd s.
 \end{align*}
Here, by partial integration, one can check that
 \begin{align*}
   &\int \ln(1-s)\,\dd s = -(1-s)\ln(1-s) - s + C_1,\\
   &\int s\ln(1-s)\,\dd s = \frac{s^2-1}{2}\ln(1-s) - \frac{s^2}{4} - \frac{s}{2} + C_2,
 \end{align*}
 where \ $C_1,C_2\in\RR$, \ which yields that
 \begin{align*}
   &\int_p^1 \ln(1-s)\,\dd s = (1-p)\ln(1-p) - 1 + p,\qquad p\in[0,1),\\
   &\int_p^1 s\ln(1-s)\,\dd s = -\frac{p^2-1}{2}\ln(1-p) + \frac{p^2}{4} + \frac{p}{2} - \frac{3}{4}, \qquad p\in[0,1).
 \end{align*}
Hence, for \ $p\in(F_X(u),1)$, \ we get
 \begin{align*}
 \ES_{X,2}(p)
    & = u+ \beta\ln(1-F_X(u))
        -\frac{2\beta}{(1-p)^2}\left(-\frac{p^2-1}{2}\ln(1-p) + \frac{p^2}{4} + \frac{p}{2} - \frac{3}{4} \right)\\
    &\phantom{=\;}  +\frac{2\beta p}{(1-p)^2} \Big((1-p)\ln(1-p) - 1 + p \Big) \\
    & = u+ \beta\ln(1-F_X(u))  - \beta\ln(1-p) + \frac{3}{2}\beta
      = \frac{3}{2}\beta + \VaR_X(p).
 \end{align*}

Note that if \ $u=0$, \ $\kappa=0$ \ and \ $\beta=1$, \ then \ $\ES_{X,2}(p)=-\ln(1-p) + \frac{3}{2}$, \ which is nothing else but the
 \ $2^{\mathrm{nd}}$-order Expected Shortfall at the level \ $p$ \ of an exponentially distributed random variable (see Example \ref{Ex_Exponencialis}).
This is expected, since if \ $u=0$, \ $\kappa=0$ \  and \ $\beta=1$, \ then \ $F_X(x) = F_{X,u}(x) = G_{0,1}(x)$, \ $0\leq x< x_{F_X}$
 with \ $x_{F_X}=\infty$, \ yielding that \ $F_X(x) = G_{0,1}(x)$, \ $x\geq 0$, \ where \ $G_{0,1}$ \ is the distribution function of an exponentially
 distributed random variable with parameter \ $1$.

Let \ $\vare\in(0,1-F_X(u))$.
\ Then the inequality
 \[
    \ES_{X,2}(1-c\vare) \leq \VaR_X(1-\vare), \qquad c\in\Big[1,\frac{1-F_X(u)}{\vare}\Big)
 \]
 takes the form
 \[
  u+ \beta\ln(1-F_X(u))  - \beta\ln(c\vare) + \frac{3}{2}\beta \leq u - \beta\ln\left(\frac{\vare}{1-F_X(u)}\right), \qquad c\in\Big[1,\frac{1-F_X(u)}{\vare}\Big).
 \]
This inequality holds if and only if \ $c\geq \ee^{\frac{3}{2}}$ \ and \ $c\in\Big[1,\frac{1-F_X(u)}{\vare}\Big)$.
\ Consequently, using that the function \ $[1,\frac{1}{\vare}]\ni c\mapsto \ES_{X,2}(1-c\vare)$ \ is monotone decreasing (see Lemma \ref{Lem_pelvehofolytonos_monoton}),
 for \ $0<\vare < (1-F_X(u))\ee^{-\frac{3}{2}}$ \ we have
 \begin{align*}
 \Pi_{\vare,2}(X)&=\inf\Big\{c \in \Big[1,\frac{1}{\vare}\Big]: \ES_{X,2}(1-c\vare)\leq \VaR_X(1-\vare) \Big\} \\
                 &= \inf\Big\{c \in \Big[1,\frac{1-F_X(u)}{\vare}\Big): c\geq \ee^{\frac{3}{2}}\Big\}
                 = \ee^{\frac{3}{2}},
 \end{align*}
 as desired.

Next, we consider the case of \ $\kappa<1$ \ with \ $\kappa\ne 0$.
\ For \ $p\in(F_X(u),1)$, \ we have
 \begin{align*}
   \ES_{X,2}(p)
    & = \frac{2}{(1-p)^2} \int_p^1(s-p)\left( u + \frac{\beta}{\kappa}\left( \left( \frac{1-s}{1-F_X(u)}\right)^{-\kappa} - 1 \right) \right)\dd s \\
    & = \frac{2}{(1-p)^2}\Big( u - \frac{\beta}{\kappa}  \Big) \int_p^1 (s-p)\,\dd s
       + \frac{2\beta(1-F_X(u))^\kappa}{(1-p)^2\kappa}\int_p^1 s(1-s)^{-\kappa}\,\dd s\\
    &\phantom{=\;}   -\frac{2\beta p(1-F_X(u))^\kappa}{(1-p)^2\kappa}\int_p^1(1-s)^{-\kappa}\,\dd s.
 \end{align*}
Here, by partial integration,
 \begin{align*}
    \int s(1-s)^{-\kappa} \,\dd s
       = -\frac{\frac{1}{\kappa}}{\left(\frac{1}{\kappa}-1\right)\left(\frac{2}{\kappa}-1\right)}
          \left( \left(\frac{1}{\kappa}-1\right)s(1-s)^{1-\kappa} + \frac{1}{\kappa} (1-s)^{1-\kappa} \right)
         + C,
 \end{align*}
 where \ $C\in\RR$.
\ Consequently, for \ $p\in(F_X(u),1)$, \ we get that
 \begin{align*}
   \ES_{X,2}(p)
      & = u- \frac{\beta}{\kappa}
        + \frac{2\beta(1-F_X(u))^\kappa}{(1-p)^2\kappa}
          \cdot\frac{\frac{1}{\kappa}}{\left(\frac{1}{\kappa}-1\right)\left(\frac{2}{\kappa}-1\right)}
          \left(\left(\frac{1}{\kappa}-1\right)p(1-p)^{1-\kappa} + \frac{1}{\kappa}(1-p)^{1-\kappa}\right)\\
    &\phantom{=\;} -\frac{2\beta p(1-F_X(u))^\kappa}{(1-p)^2\kappa}\cdot\frac{(1-p)^{1-\kappa}}{1-\kappa}\\
    & = u- \frac{\beta}{\kappa}
        +\frac{2\beta(1-F_X(u))^\kappa}{\kappa(1-\kappa)(2-\kappa)}
          (1-p)^{-\kappa}\\
    & = \VaR_X(p) + \frac{\beta(3-\kappa)}{(1-\kappa)(2-\kappa)} \left(\frac{1-p}{1-F_X(u)}\right)^{-\kappa}.
 \end{align*}
Let \ $\vare\in(0,1-F_X(u))$.
\ Then the inequality
 \[
    \ES_{X,2}(1-c\vare) \leq \VaR_X(1-\vare), \qquad c\in\Big[1,\frac{1-F_X(u)}{\vare}\Big)
 \]
 takes the form
 \[
   u- \frac{\beta}{\kappa}
        +\frac{2\beta(1-F_X(u))^\kappa}{\kappa(1-\kappa)(2-\kappa)}
          (c\vare)^{-\kappa}
     \leq u + \frac{\beta}{\kappa}\left( \left( \frac{\vare}{1-F_X(u)}\right)^{-\kappa}  - 1 \right) , \qquad c\in\Big[1,\frac{1-F_X(u)}{\vare}\Big).
 \]
This inequality holds if and only if
 \[
   c\geq \left(\frac{2}{(1-\kappa)(2-\kappa)}\right)^{\frac{1}{\kappa}} \qquad  \text{and} \qquad c\in\Big[1,\frac{1-F_X(u)}{\vare}\Big).
 \]
Consequently, using that the function \ $[1,\frac{1}{\vare}]\ni c\mapsto \ES_{X,2}(1-c\vare)$ \ is monotone decreasing (see Lemma \ref{Lem_pelvehofolytonos_monoton}),
 for \ $0<\vare < (1-F_X(u))\big((1-\kappa)(2-\kappa)/2\big)^{\frac{1}{\kappa}}$ \ we have
 \begin{align*}
 \Pi_{\vare,2}(X)&=\inf\Big\{c \in \Big[1,\frac{1}{\vare}\Big]: \ES_{X,2}(1-c\vare)\leq \VaR_X(1-\vare) \Big\} \\
                 &= \inf\Big\{c \in \Big[1,\frac{1-F_X(u)}{\vare}\Big): c\geq  \left(\frac{2}{(1-\kappa)(2-\kappa)}\right)^{\frac{1}{\kappa}} \Bigg\}
                  = \left(\frac{2}{(1-\kappa)(2-\kappa)}\right)^{\frac{1}{\kappa}},
 \end{align*}
 as desired.

Now, we turn to prove \eqref{help_ES2_VAR_ratio}.
In case of \ $\kappa=0$, \ we have \ $\lim_{p\uparrow 1} \VaR_X(p)=\infty$ \ and
 \[
   \lim_{p\uparrow 1} \frac{\ES_{X,2}(p)}{\VaR_X(p)}
      =  \lim_{p\uparrow 1} \frac{\VaR_X(p) + \frac{3}{2}\beta}{\VaR_X(p)}
      = 1.
 \]
In case of \ $\kappa\in(0,1)$, \ we have \ $\lim_{p\uparrow 1} \VaR_X(p)=\infty$ \ and
 \begin{align*}
   \lim_{p\uparrow 1} \frac{\ES_{X,2}(p)}{\VaR_X(p)}
   & = \lim_{p\uparrow 1} \frac{\VaR_X(p) + \frac{\beta(3-\kappa)}{(1-\kappa)(2-\kappa)} \left(\frac{1-p}{1-F_X(u)}\right)^{-\kappa}}{\VaR_X(p)} \\
   & = 1 + \lim_{p\uparrow 1} \frac{\frac{\beta(3-\kappa)}{(1-\kappa)(2-\kappa)} \left(\frac{1-p}{1-F_X(u)}\right)^{-\kappa}}
                                  {u + \frac{\beta}{\kappa}\left( \left( \frac{1-p}{1-F_X(u)}\right)^{-\kappa} - 1 \right)}\\
   & = 1 + \frac{\frac{\beta(3-\kappa)}{(1-\kappa)(2-\kappa)} }{ \frac{\beta}{\kappa} }
     = \frac{2}{(1-\kappa)(2-\kappa)}.
 \end{align*}
In case of \ $\kappa<0$, \ we have \ $\lim_{p\uparrow 1} \VaR_X(p)=u-\frac{\beta}{\kappa}$ \ and
 \begin{align*}
   \lim_{p\uparrow 1} \frac{\ES_{X,2}(p)}{\VaR_X(p)}
    = 1 + \lim_{p\uparrow 1} \frac{\frac{\beta(3-\kappa)}{(1-\kappa)(2-\kappa)} \left(\frac{1-p}{1-F_X(u)}\right)^{-\kappa}}
                                   {\VaR_X(p)}
    = 1 + \frac{0}{u-\frac{\beta}{\kappa}}
     = 1,
 \end{align*}
 as desired.
\proofend

\section{$\PELVES$ \ of regularly varying distributions}\label{Sec_PELVE2_reg_var}

First, we recall the notion of regularly varying functions.

\begin{Def}\label{Def_reg_var}
A measurable function \ $U: (x_0,\infty) \to (0,\infty)$ \ (where \ $x_0\geq 0$) \ is called regularly varying at infinity with
 index \ $\rho \in \RR$ \ if for all \ $q>0$, \ we have
 \[
   \lim_{x\to\infty} \frac{U(qx)}{U(x)} = q^\rho .
 \]
In case of \ $\rho = 0$, \ we call \ $U$ \ slowly varying at infinity.
A measurable function \ $V: (0,x_0) \to (0,\infty)$ \ (where \ $x_0>0$) \ is called regularly varying at \ $0$ \ with
 index \ $\kappa \in \RR$ \ if for all \ $q>0$, \ we have
 \[
   \lim_{x\downarrow 0} \frac{V(qx)}{V(x)} = q^\kappa.
 \]
\end{Def}

Next, we recall the notion of regularly varying non-negative random variables.

\begin{Def}
A non-negative random variable \ $X$ \ is called regularly varying with index \ $\alpha\geq 0$ \
 if \ $\overline F_X(x) := \PP(X > x) > 0$ \ for all \ $x > 0$, \ and \ $\overline F_X$ \ is regularly varying at infinity with index \ $-\alpha$.
\end{Def}

Next, we study the asymptotic behaviour
 of the \ $\PELVES$ \ of \ $X$ \ at a level \ $\vare$ \ tending to \ $0$.
Recall that if \ $X$ \ is a regularly varying non-negative random variable with index \ $\alpha>1$, \ then \ $\EE(X)<\infty$, \ i.e., \ $X\in L^1$,
 \ and hence in this case \ $\PELVES$ \ of \ $X$ \ at any level \ $\vare\in(0,1)$ \ is well-defined.

\begin{Thm}\label{Thm_regularly_varying}
Let \ $X$ \ be a regularly varying non-negative random variable with index \ $\alpha>1$ \ such that \ $(0,1)\ni p \mapsto \VaR_X(p)$ \ is continuous.
Then
 \[
    \lim_{\vare\downarrow 0} \Pi_{\vare,2}(X) = \left( \frac{2\alpha^2}{(\alpha-1)(2\alpha-1)} \right)^\alpha.
 \]
\end{Thm}

\noindent{\bf Proof.}
Recall that
 \begin{align}\label{VaR_asymptotic}
   \lim_{\vare\downarrow 0}
         \frac{\VaR_X(1-t\vare)}{\VaR_X(1-\vare)}
       = t^{-\frac{1}{\alpha}},
     \qquad t>0,
 \end{align}
 i.e., the function \ $(0,1)\ni p \mapsto \VaR_X(p)$ \ is regularly varying at \ $0$ \ with index \ $-\frac{1}{\alpha}$, \
 see, e.g., Li and Wang \cite[formula (A.9)]{LiWan}.
For completeness, we present a proof of \eqref{VaR_asymptotic}.
Since \ $X$ \ is regularly varying with index \ $\alpha>1$, \ we have \ $\frac{1}{\overline F_X}$ \ is a monotone increasing and regularly varying
 function at infinity with index \ $\alpha$ \ satisfying \ $\lim_{x\to\infty} \frac{1}{\overline F_X(x)}=\infty$. \
Consequently, by Resnick \cite[Proposition 0.8/(v)]{Res},
 \ $\big(1/\overline F_X\big)^{-1}$ \ is a regularly varying function at infinity with index \ $\frac{1}{\alpha}$, \ where
 \begin{align*}
    \left(\frac{1}{\overline F_X}\right)^{-1}(x) & :=\inf\left\{ y\in\RR : \frac{1}{\overline F_X(y)} \geq x\right\}
            = \inf\left\{ y\in\RR : \frac{1}{x} \geq \overline F_X(y) \right\} \\
           & = \inf\left\{ y\in\RR : F_X(y) \geq 1 - \frac{1}{x} \right\}
             = \VaR_X\Big(1-\frac{1}{x}\Big),  \qquad x>0.
 \end{align*}
Hence
 \begin{align*}
   t^{\frac{1}{\alpha}}
      = \lim_{x\to\infty}
         \frac{ \left(\frac{1}{\overline F_X}\right)^{-1}(tx)}
              {  \left(\frac{1}{\overline F_X}\right)^{-1}(x) }
      = \lim_{x\to\infty}
         \frac{\VaR_X\Big(1-\frac{1}{tx}\Big)}{\VaR_X\Big(1-\frac{1}{x}\Big)}
      = \lim_{\vare\downarrow 0}
         \frac{\VaR_X\Big(1-\frac{1}{t}\vare\Big)}{\VaR_X\Big(1-\vare\Big)},
     \qquad t>0,
 \end{align*}
 and consequently we have \eqref{VaR_asymptotic}, as desired.

Recall also that
  \begin{align}\label{ES_asymptotic}
   \lim_{\vare\downarrow 0}
         \frac{\ES_X(1-\vare)}{\VaR_X(1-\vare)}
       = \frac{\alpha}{\alpha-1},
     \qquad t>0,
 \end{align}
 see, e.g., Li and Wang \cite[formula (A.10)]{LiWan}.
For completeness, we present a proof of \eqref{ES_asymptotic}.
For each \ $\vare\in(0,1)$, \ we have
 \begin{align*}
   \frac{\ES_X(1-\vare)}{\VaR_X(1-\vare)}
    = \frac{\frac{1}{\vare}\int_{1-\vare}^1 \VaR_X(u)\,\dd u }{\VaR_X(1-\vare)}
    = \frac{\int_0^{\vare} \VaR_X(1-v)\,\dd v }{\vare\VaR_X(1-\vare)}.
 \end{align*}
By \eqref{VaR_asymptotic}, the function \ $(0,1)\ni v\mapsto \VaR_X(1-v)$ \ is regularly varying at \ $0$ \ with index \ $-\frac{1}{\alpha}>-1$,
 \ and hence, by Karamata theorem for regularly varying functions at zero (see Lemma \ref{Lem_Karamata_zero}), we have
 \[
   \lim_{\vare\downarrow 0}
      \frac{\int_0^{\vare} \VaR_X(1-v)\,\dd v }{\vare\VaR_X(1-\vare)}
    = \frac{1}{-\frac{1}{\alpha}+1}
    =   \frac{\alpha}{\alpha - 1},
 \]
 yielding \eqref{ES_asymptotic}.

Next, we prove that
 \[
    \lim_{\vare\downarrow 0}
       \frac{\ES_{X,2}(1-\vare)}{\VaR_X(1-\vare)}
    = \frac{2\alpha^2}{(\alpha-1)(2\alpha-1)}.
 \]
For each \ $\vare\in(0,1)$, \ we have
 \begin{align*}
  \frac{\ES_{X,2}(1-\vare)}{\VaR_X(1-\vare)}
   & = \frac{\frac{2}{\vare^2} \int_{1-\vare}^1 (u - (1-\vare)) \VaR_X(u)\,\dd u }
           {\VaR_X(1-\vare)}
     = \frac{2 \int_0^\vare (\vare - v) \VaR_X(1-v)\,\dd v }{\vare^2 \VaR_X(1-\vare)} \\
   & = 2\frac{ \int_0^\vare \VaR_X(1-v)\,\dd v }{\vare \VaR_X(1-\vare)}
       - 2\frac{\int_0^\vare v \VaR_X(1-v)\,\dd v }{\vare\cdot \vare\VaR_X(1-\vare)}\\
   & = 2\frac{\ES_X(1-\vare)}{\VaR_X(1-\vare)} - 2\frac{\int_0^\vare v \VaR_X(1-v)\,\dd v }{\vare\cdot \vare\VaR_X(1-\vare)}.
 \end{align*}
Here the function \ $(0,1)\ni v\mapsto v \VaR_X(1-v)$ \ is regularly varying at \ $0$ \ with index \ $\frac{\alpha-1}{\alpha}>-1$, \
 since, using \eqref{VaR_asymptotic}, we have
 \[
   \lim_{v\downarrow 0} \frac{tv\VaR_X(1-tv)}{v\VaR_X(1-v)}
     = t\cdot t^{-\frac{1}{\alpha}} = t^{\frac{\alpha-1}{\alpha}}, \qquad t>0.
 \]
So, by Karamata theorem for regularly varying functions at zero (see Lemma \ref{Lem_Karamata_zero}), we have
 \[
   \lim_{\vare\downarrow 0}
   \frac{\int_0^\vare v \VaR_X(1-v)\,\dd v }{\vare\cdot \vare\VaR_X(1-\vare)}
     = \frac{1}{\frac{\alpha-1}{\alpha}+1}
     = \frac{\alpha}{2\alpha-1}.
 \]
Hence, using \eqref{ES_asymptotic}, we get
 \[
 \lim_{\vare\downarrow 0}
       \frac{\ES_{X,2}(1-\vare)}{\VaR_X(1-\vare)}
    = \frac{2\alpha}{\alpha-1} - \frac{2\alpha}{2\alpha-1}
    =  \frac{2\alpha^2}{(\alpha-1)(2\alpha-1)}.
 \]
Consequently, using \eqref{VaR_asymptotic}, we have
 \begin{align}\label{help_ES_9}
  \begin{split}
     \lim_{\vare\downarrow 0}
        \frac{\ES_{X,2}(1-t\vare)}{\VaR_X(1-\vare)}
     = \lim_{\vare\downarrow 0}
        \frac{\ES_{X,2}(1-t\vare)}{\VaR_X(1-t\vare)}
        \cdot \frac{\VaR_X(1-t\vare)}{\VaR_X(1-\vare)}
     = \frac{2\alpha^2}{(\alpha-1)(2\alpha-1)} t^{-\frac{1}{\alpha}},
     \qquad t>0.
  \end{split}
 \end{align}
Note that for each \ $\alpha>1$, \ the function \ $(0,\infty)\ni t\mapsto \frac{2\alpha^2}{(\alpha-1)(2\alpha-1)} t^{-\frac{1}{\alpha}}$ \
 is strictly monotone decreasing, and it takes value \ $1$ \ if and only if
 \[
     t=\left(\frac{2\alpha^2}{(\alpha-1)(2\alpha-1)}\right)^\alpha .
 \]
Let \ $t_1$ \ and \ $t_2$ \ be such that
 \[
    0< t_1 < \left( \frac{2\alpha^2}{(\alpha-1)(2\alpha-1)}\right)^\alpha < t_2,
 \]
 yielding
 \[
   \frac{2\alpha^2}{(\alpha-1)(2\alpha-1)} t_2^{-\frac{1}{\alpha}}
     < 1 < \frac{2\alpha^2}{(\alpha-1)(2\alpha-1)} t_1^{-\frac{1}{\alpha}}.
 \]
Hence, using \eqref{help_ES_9}, for sufficiently small \ $\vare>0$ \ (which may depend on \ $t_1$ \ and \ $t_2$),
 \ we have
 \[
   \frac{\ES_{X,2}(1-t_2\vare)}{\VaR_X(1-\vare)}< 1
   \qquad \text{and}\qquad
   \frac{\ES_{X,2}(1-t_1\vare)}{\VaR_X(1-\vare)}> 1.
 \]
Since \ $X$ \ is non-negative and regularly varying, we have \ $\VaR_X(1-v)\to\infty$ \ as \ $v\downarrow 0$, \ and hence for sufficiently small \ $\vare>0$,
 \ we get \ $\VaR_X(1-\vare)>0$ \ and
 \begin{align}\label{help_ES_10}
   \ES_{X,2}(1-t_2\vare) < \VaR_X(1-\vare) < \ES_{X,2}(1-t_1\vare).
 \end{align}

Using (again) \ $\VaR_X(1-v)\to\infty$ \ as \ $v\downarrow 0$, \ and \ $\ES_{X,2}(0)<\infty$, \ we have \ $\ES_{X,2}(0) < \VaR_X(1-\vare)$ \
 for sufficiently small \ $\vare>0$.
\ Hence, using Proposition \ref{Pro1}, for sufficiently small \ $\vare>0$, \ we have
 \ $\Pi_{\vare,2}(X)\in[1,\frac{1}{\vare}]$ \ is a solution of the equation
 \ $\ES_{X,2}(1-c\vare) = \VaR_X(1-\vare)$, \ $c\in[1,\frac{1}{\vare}]$.
\ Consequently, using \eqref{help_ES_10}, the definition of \ $\PELVES$, \
 and the continuity and monotone increasing property of \ $\ES_{X,2}$ \ (see Lemma \ref{Lem_pelvehofolytonos_monoton}),
 we get \ $\Pi_{\vare,2}(X)\in(t_1,t_2]$ \ for sufficiently small \ $\vare>0$.
\ Since \ $(t_1,t_2]$ \ can be chosen as a neighbourhood of
 \ $\big(2\alpha^2/((\alpha-1)(2\alpha-1))\big)^\alpha$ with arbitrarily small length, the statement follows.
\proofend

\begin{Rem}
In Theorem \ref{Thm_regularly_varying}, the limit is nothing else but the \ $\PELVES$ \ of a Pareto distributed random variable with parameters \ $k>0$
 \ and \ $\alpha>1$ \ at any level less then or equal to \ $\Big(\frac{(\alpha-1)(2\alpha-1)}{2\alpha^2}\Big)^\alpha$, \
  see Example \ref{Ex_Pareto}.
So Theorem \ref{Thm_regularly_varying} is in accordance with Theorem 3 in Li and Wang \cite{LiWan}.
Note that a Pareto distributed random variable with parameters \ $k>0$ \ and \ $\alpha>0$ \ is regularly varying with
 index \ $-\alpha$, \ and a random variable with generalized Pareto distribution having
parameters \ $\kappa>0$ \ and \ $\beta>0$ \ (see, Definition \ref{Def_Gen_Par})
 is regularly varying with index \ ${ -\frac{1}{\kappa}}$.
\proofend
\end{Rem}

\section{Simulations and real data analysis for \ $\PELVE_2$}\label{Sec_sim}

First, we present an empirical estimator of \ $\PELVE_2$ \ of a random variable.
Let \ $X\in L^1$, \ $m\in\NN$, \ and let \ $X_1,\ldots,X_m$ \ be  independent
 and identically distributed random variables such that their common distribution coincides with that of \ $X$,
  i.e., $X_1,\ldots,X_m$ \ is a sample of length $m$ for $X$.
\ Let \ $X_1^*\leq X_2^* \leq \ldots\leq X_m^*$ \ be the corresponding ordered sample.
Given \ $p\in(0,1)$, \ an empirical estimator of \ $\VaR_X(p)$ \ based on \ $X_1,\ldots,X_m$ \ is given by
 \[
  \widehat{\VaR_X}(p):=X_i^*\qquad \text{if \ $p\in\left(\frac{i-1}{m},\frac{i}{m}\right]$, \ $i=1,\ldots,m$.}
 \]
Following Acerbi \cite[Section 7]{Ace} (where one can find a construction of empirical estimators of spectral risk measures),
 given \ $p\in[0,1)$, \ an empirical estimator of the $2^{\mathrm{nd}}$-order Expected Shortfall
 \ $\ES_{X,2}(p)$ \ based on \ $X_1,\ldots,X_m$ \ is given by a weighted sum of $X_1^*,\ldots,X_m^*$:
 \[
   \widehat{\ES_{X,2}}(p):=\sum_{i=1}^m w_i X_i^*,
 \]
 where
 \[
   w_i:=\int_{(i-1)/m}^{i/m} \frac{2(s-p)}{(1-p)^2}\bbone_{[p,1)}(s)\,\dd s,\qquad i=1,\ldots,m.
 \]
We check that if \ $p\in[0,\frac{m-1}{m})$, \ then
 \begin{align}\label{help_suly_1}
   w_i = \begin{cases}
           0 & \text{if \ $i\leq \lfloor mp\rfloor$,}\\[1mm]
           \frac{1}{(1-p)^2} \left(\frac{i}{m}-p\right)^2 & \text{if \ $i= \lfloor mp\rfloor+1$,}\\[1mm]
           \frac{1}{(1-p)^2} \left( \left(\frac{i}{m}-p\right)^2 - \left(\frac{i-1}{m}-p\right)^2  \right)
                                                           & \text{if \ $\lfloor mp\rfloor+2\leq i\leq m$,}
         \end{cases}
 \end{align}
 and if \ $p\in[\frac{m-1}{m},1)$, \ then
 \begin{align}
   w_i= \begin{cases}\label{help_suly_2}
           0 & \text{if \ $i=1,\ldots,m-1$,}\\
           1 & \text{if \ $i=m$.}
         \end{cases}
 \end{align}
If \ $p\in[0,\frac{1}{m})$, \ then \ $\lfloor mp\rfloor=0$ \ and
 \[
  w_1=\int_{p}^{1/m} \frac{2(s-p)}{(1-p)^2}\,\dd s = \frac{1}{(1-p)^2} \left(\frac{1}{m}-p\right)^2,
 \]
 and
 \[
  w_i=\int_{(i-1)/m}^{i/m} \frac{2(s-p)}{(1-p)^2}\,\dd s
     = \frac{1}{(1-p)^2} \left( \left(\frac{i}{m}-p\right)^2 - \left(\frac{i-1}{m}-p\right)^2  \right) ,
     \qquad i=2,\ldots,m,
 \]
 yielding \eqref{help_suly_1} in case of \ $p\in[0,\frac{1}{m})$.
If \ $p\in[\frac{1}{m},\frac{2}{m})$, \ then \ $\lfloor mp\rfloor=1$, \ $w_1=\int_0^{1/m} 0\,\dd s=0$, \ and
 \[
  w_2=\int_{p}^{2/m} \frac{2(s-p)}{(1-p)^2}\,\dd s = \frac{1}{(1-p)^2} \left(\frac{2}{m}-p\right)^2,
 \]
 and
 \[
  w_i=\int_{(i-1)/m}^{i/m} \frac{2(s-p)}{(1-p)^2}\,\dd s
     = \frac{1}{(1-p)^2} \left( \left(\frac{i}{m}-p\right)^2 - \left(\frac{i-1}{m}-p\right)^2  \right) ,
     \qquad i=3,\ldots,m,
 \]
 yielding \eqref{help_suly_1} in case of \ $p\in[\frac{1}{m},\frac{2}{m})$.
The case \ $p\in[\frac{2}{m},\frac{m-1}{m})$ \ can be handled similarly.
If \ $p\in[\frac{m-1}{m},1)$, \ then \ $\lfloor mp\rfloor=m-1$, \ $w_i=0$, \ $i=1,\ldots,m-1$, \ and
 \[
  w_m=\int_{p}^{1} \frac{2(s-p)}{(1-p)^2}\,\dd s =1,
 \]
 yielding \eqref{help_suly_1} in case of \ $p\in[\frac{m-1}{m},1)$, \ as desired.

Consequently, if \ $p\in[0,\frac{m-1}{m})$, \ then
 \[
   \widehat{\ES_{X,2}}(p)
     = \frac{1}{(1-p)^2} \!\!\left[ \!\left(\frac{\lfloor mp\rfloor+1}{m}-p\right)^2 X_{\lfloor mp\rfloor+1}^*
                                 + \sum_{i=\lfloor mp\rfloor+2}^m\!
                                     \left(\left(\frac{i}{m}-p\right)^2 -  \left(\frac{i-1}{m}-p\right)^2\right)X_i^*
                          \right],
 \]
 and if \ $p\in[\frac{m-1}{m},1)$, \ then
 \[
   \widehat{\ES_{X,2}}(p) = X_m^*.
 \]

Given \ $\vare\in(0,1)$, \ an empirical estimator of the \ $\PELVE_2$ \ value \ $\Pi_{\vare,2}(X)$ \ based on \ $X_1,\ldots,X_m$ \ can be defined as
 \begin{align}\label{help_PELVE2_emp_est}
  \widehat{\Pi_{\vare,2}(X)}:=\inf\Big\{c \in \Big[1,\frac{1}{\vare}\Big]: \widehat{\ES_{X,2}}(1-c\vare)\leq \widehat{\VaR_X}(1-\vare) \Big\},
 \end{align}
 where \ $\inf\emptyset=\infty$.

In principle, the empirical estimator \eqref{help_PELVE2_emp_est} of the \ $\PELVE_2$ \ value \ $\Pi_{\vare,2}(X)$ \
 can be used even if the random variables \ $X_1,\ldots,X_m$ \ are not independent or identically distributed.
We will do so in analyzing real data.

\begin{figure}[ht]
 \centering
 \includegraphics[width=6cm]{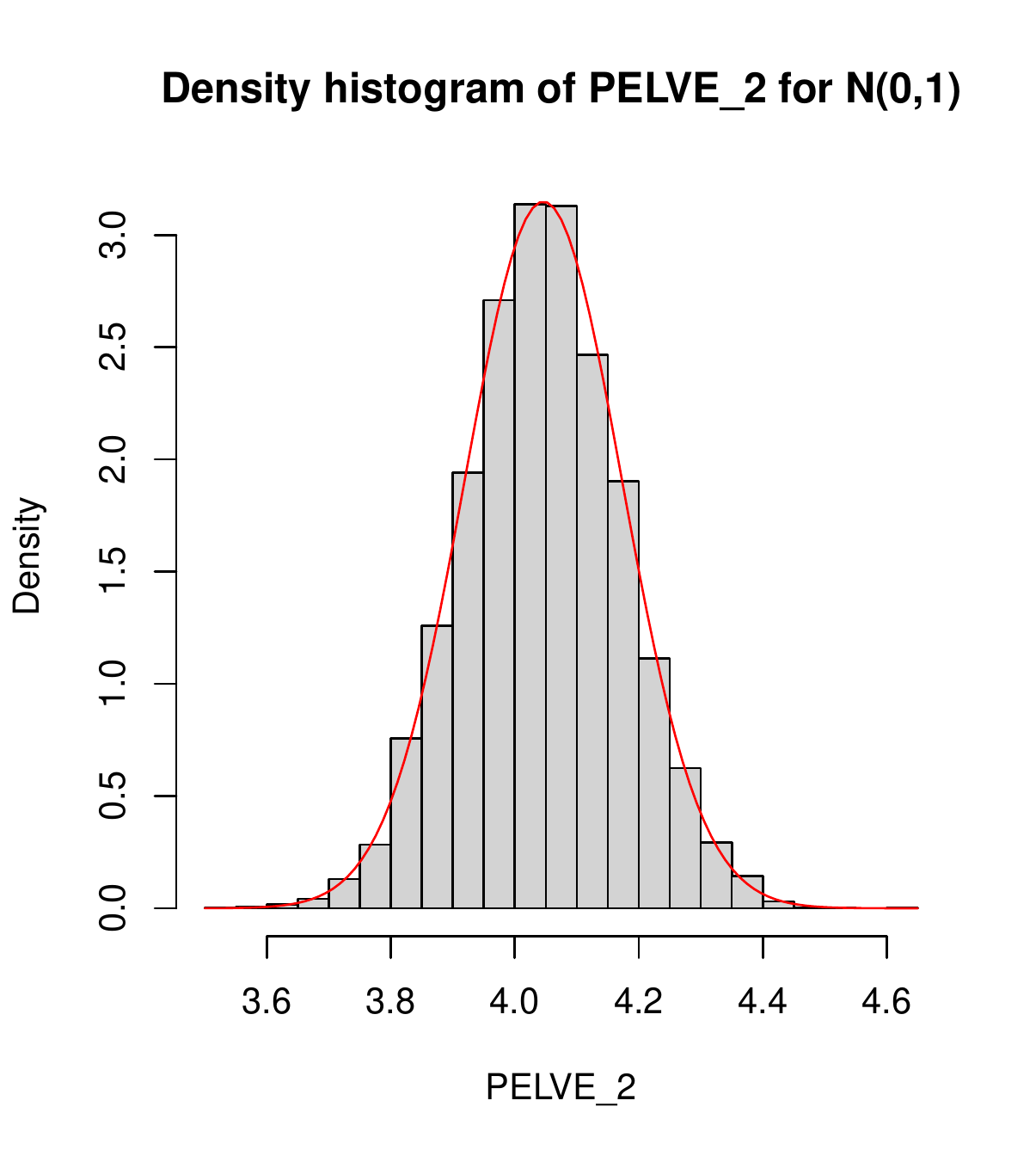}
 \caption{Density histogram of empirical $\PELVE_2$ values of a standard normal distribution at level $\vare=0.05$
          based on $10000$ samples of length $m=5000$. The red curve is the density function of the fitted normal distribution.}
 \label{Fig_norm_sim}
\end{figure}

For illustrative purposes, we present a simulation result for calculating the $\PELVE_2$ value of a standard normally distributed random variable
 at the level \ $\vare=0.05$.
\ We generated \ $10000$ \ samples of length $m=5000$ for a standard normally distributed random variable \ $X$.
For each generated sample, we calculated the empirical estimator $\widehat{\Pi_{\vare,2}(X)}$ \ (given in \eqref{help_PELVE2_emp_est})
 of the $\PELVE_2$ value of \ $X$ \ at the given level \ $\vare$.
\ Then we made a density histogram based on the $10000$ estimated $\PELVE_2$ values, see Figure \ref{Fig_norm_sim}.
On this figure, we also plotted the density function of the fitted normally distribution in red.
The theoretical $\PELVE_2$ value of a standard normally distributed random variable at the level $\vare=0.05$ is approximately $4.040815$.
The sample mean of the 10000 estimated $\PELVE_2$ values is \ $4.046066$, \ which is quite close to the theoretical value.
Based on Figure \ref{Fig_norm_sim}, one could conjecture that a kind of central limit theorem might hold for $\PELVE_2$ in case of
 a standard normal distribution.
We do not study this question here.
We only note that in case of $\PELVE$, such a result is available due to Li and Wang \cite[Theorem 4]{LiWan}.

\begin{figure}[ht!]
\centering
\begin{subfigure}[t]{0.4\textwidth}
\centering
\includegraphics[width=\textwidth,height=6cm]{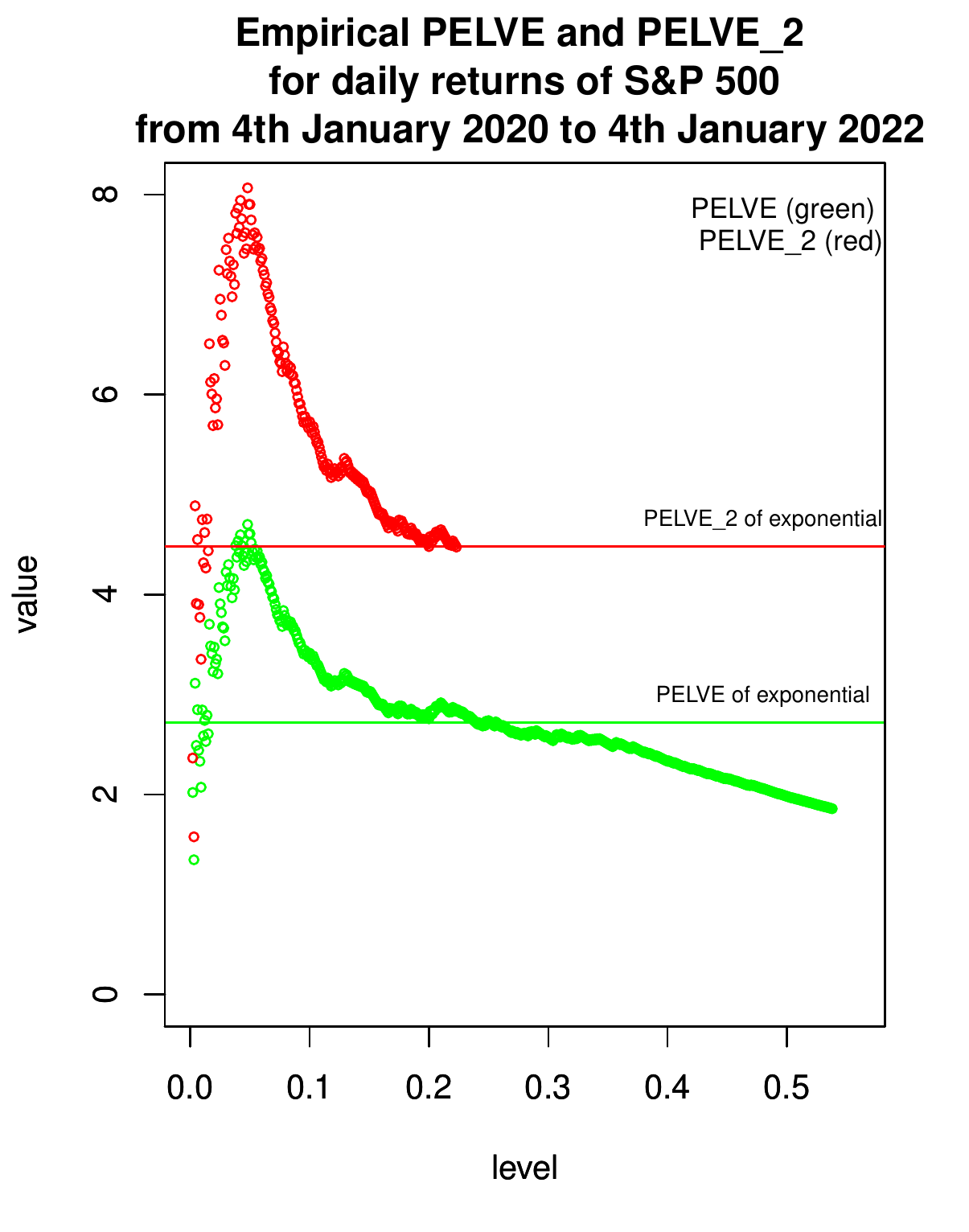}
\caption{Case (i).}
\label{Fig_SP500_4}
\end{subfigure}
\hspace{2cm}
\begin{subfigure}[t]{0.4\textwidth}
\centering
\includegraphics[width=\textwidth,height=6cm]{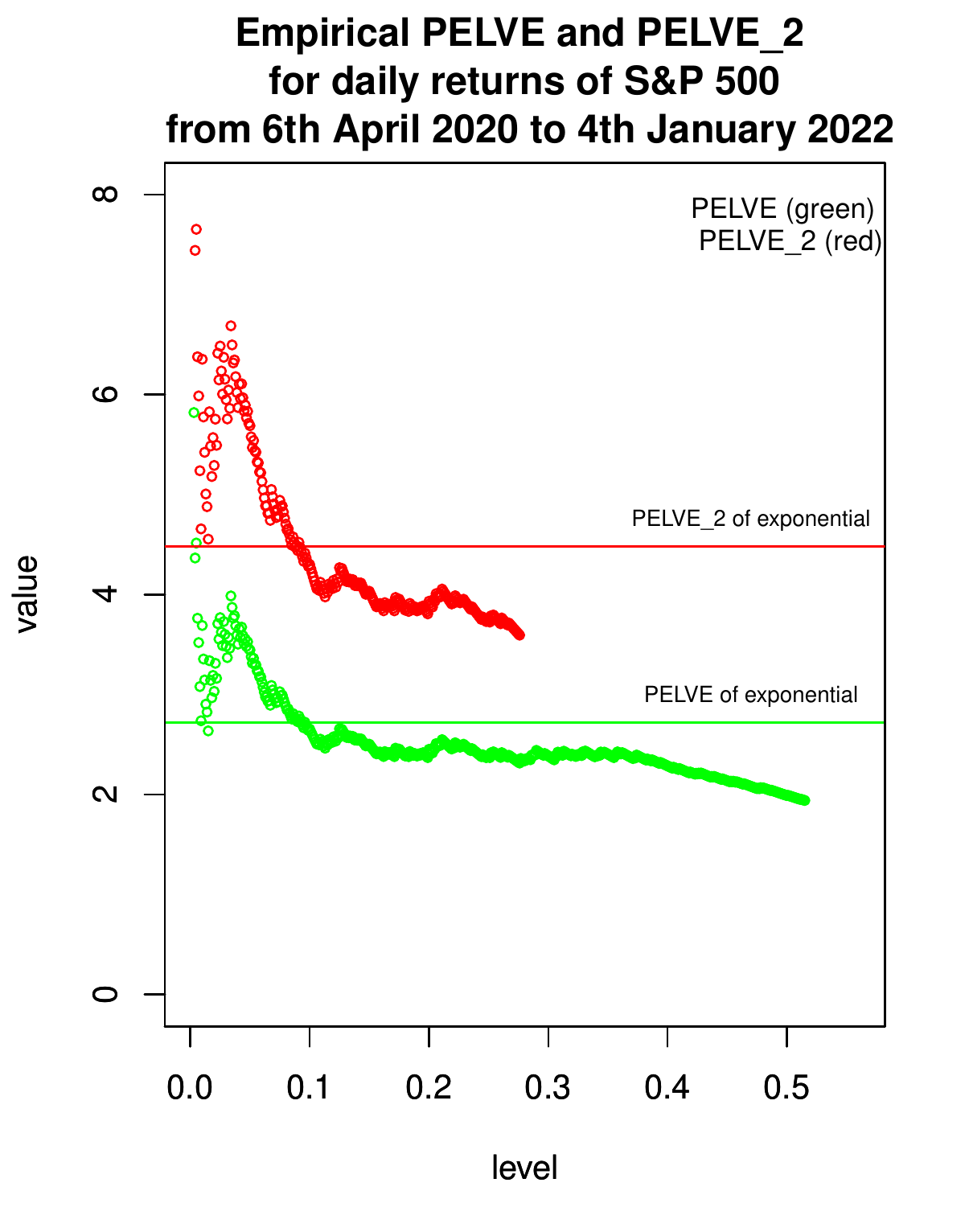}
\caption{Case (ii).}
\label{Fig_SP500_5}
\end{subfigure}
\caption{Empirical $\PELVE$ and $\PELVE_2$ values for daily returns of S\&P 500.}
\label{Fig_real}
\end{figure}

As real data applications, we calculate the empirical estimator of \ $\PELVE_2$ \ for S\&P 500 daily returns
 based on two data sets:
 (i) ranging from 4th January 2020 to 4th January 2022, and (ii) ranging from 6th April 2020 to 4th January 2022.
The S\&P 500 historical data sets were downloaded from Investing.com.
Note that in case (i) the data set contains approximately four months before the COVID-19 crisis started in Europe
 (i.e., before April 2020), and in case (ii) the data set just starts when the COVID-19 crisis started in Europe.
Recall that, given some asset prices \ $S_t$, \ $t=0,1,\ldots,N$, \ where \ $N\in\NN$, \ the one-period
 (linear) return at time \ $t=1,\ldots,N$ \ is defined by \ $S_t/S_{t-1}-1$.
In the used data sets, the daily returns are rounded off to two decimal places.
For both data sets in question, we calculated the empirical \ $\PELVE_2$ estimator \eqref{help_PELVE2_emp_est}
 and the empirical $\PELVE$ estimator at levels ranging from $0.001$ to $0.56$, \ see Figure \ref{Fig_real}.
The empirical $\PELVE$ estimator is not presented in the present paper,
 we used the same empirical estimators as Li and Wang \cite[Section 5]{LiWan} and Fiori and Rosazza Gianin \cite[Section 5]{FioGia}.

On Figure \ref{Fig_real}, the red horizontal lines correspond to the $\PELVE_2$ value of an exponential distribution,
 i.e., \ $\ee^{3/2}\approx 4.482$ \ (see Example \ref{Ex_Exponencialis});
 and the green horizontal lines correspond to the \ $\PELVE$ value of an exponential distribution, i.e., \ $\ee\approx 2.718$
 \ (see Li and Wang \cite[Example 5, part (ii)]{LiWan}).
On Figure \ref{Fig_real}, one can see that there are levels for which no empirical $\PELVE_2$ or $\PELVE$ values are plotted.
It just means that the corresponding empirical $\PELVE_2$ or $\PELVE$ values are infinity.
For example, on the right subfigure of Figure \ref{Fig_real}, no $\PELVE_2$ values are plotted at levels greater than (approximately) \ $0.27$.
As shown in the left subfigure of Figure \ref{Fig_real}, most of the empirical $\PELVE_2$ values are above \ $\ee^{3/2}$, \
 but it is not the case for the right subfigure of Figure \ref{Fig_real}.
A possible explanation for this phenomenon is that the data set used for the left subfigure of Figure \ref{Fig_real} contains four months daily returns
 of S\&P 500 before the start of the COVID-19 crisis in Europe (i.e., before April 2020),
 while the data set used for the right subfigure of Figure \ref{Fig_real} does not include these four months, it just starts at April 2020.
Our real data applications may suggest that $\PELVE_2$ might be an indicator for structural changes in stock prices.
Note also that the empirical $\PELVE_2$ values on the left subfigure of Figure \ref{Fig_real} are greater than the corresponding empirical $\PELVE_2$ values
 on the right subfigure of {Figure \ref{Fig_real}.

Finally, we present another approach to illustrate the changes in the $\PELVE$ and $\PELVE_2$ values caused by COVID-19
based on a S\&P 500 historical data set.
Fixing the level $\vare=0.05$, for each day starting from May 27, 2003 and ending at July 28, 2022,
 we calculate a corresponding $\PELVE$ and $\PELVE_2$ value based on the previous 99 days and the day in question itself
 (altogether 100 days).
The graphs are shown in Figure \ref{PELVE}, where it is visible that both risk measures had a high peak
at around March 2020 (the approximate starting date of COVID-19 crisis in Europe) as the effect of the pandemic.
Note also that the \ $\PELVE_2$ \ shows this effect more significantly.
 \begin{figure}[ht!]
\centering
\includegraphics[width=\textwidth, height=8cm]{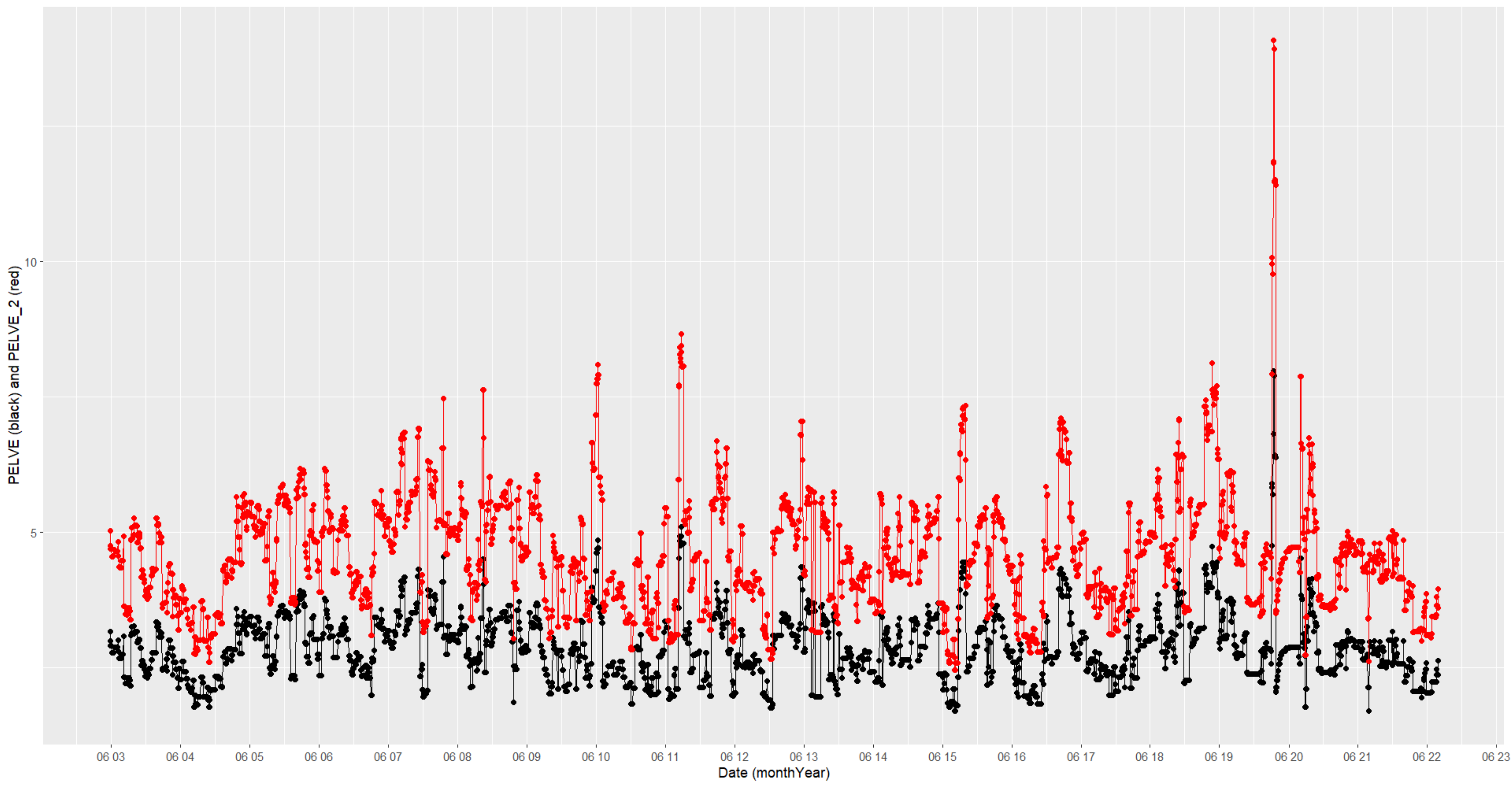}
\caption{$\PELVE$ (black) and $\PELVE_2$ (red) values.}
\label{PELVE}
\end{figure}

We used the open software R for making the simulations and real data analysis.

\appendix

\vspace*{5mm}

\noindent{\bf\Large Appendices}

\section{Some properties of higher-order Expected Shortfalls}\label{App_2ndES_prop}

This appendix is devoted to study some properties of higher-order Expected Shortfalls given in Definition \ref{Def_ES_n}
 such as finiteness, continuity, monotonicity, additivity for comonotonic random variables and connection with weak convergence.
These results generalize the corresponding known properties for (the usual, i.e., first order) Expected Shortfall.

Our first result states the finiteness of higher-order Expected Shortfalls of random variables in \ $L^1$.

\begin{Lem}\label{Lem_ESn_vegesseg}
For each \ $X\in L^1$, \ $n\in\NN$, \ and \ $p\in[0,1)$, \ we have \ $\ES_{X,n}(p)\in\RR$.
\end{Lem}

\noindent{\bf Proof.}
Let us define \ $h_p:[0,1]\to [0,1]$, \
 \begin{align}\label{Def_hp}
   h_p(s):=\left(\frac{s-p}{1-p}\right)^n\bone_{[p,1]}(s),\qquad s\in[0,1].
 \end{align}
Then \ $h_p$ \ is a distortion function in the sense of Dhaene et al.\ \cite[Definition 2]{DhaKukLinTan}, i.e.,
 \ $h_p:[0,1]\to[0,1]$ \ is nondecreasing such that \ $h_p(0) = 0$ \ and \ $h_p(1) = 1$.
\ Further, using formula (11) in Dhaene et al.\ \cite{DhaKukLinTan}, we have
 \[
   \ES_{X,n}(p)=\int_0^1 \VaR_X(s)\,\dd h_p(s)
               =\int_0^1 \VaR_X(1-s)\,\dd \overline h_p(s),
               \qquad p\in[0,1),
 \]
 where \ $\overline h_p:[0,1]\to[0,1]$, \  $\overline h_p(s):=1-h_p(1-s)$, \ $s\in[0,1]$,
 \ is also a distortion function.
Then, using formula (3) and Theorem 6 in Dhaene et al.\ \cite{DhaKukLinTan}, we get that
 \begin{align}\label{formula_ESn}
  \begin{split}
   \ES_{X,n}(p)
   & = -\int_{-\infty}^0 (1-\overline h_p(\overline F_X(x)))\,\dd x
      + \int_0^\infty \overline h_p(\overline F_X(x))\,\dd x \\
   & = -\int_{-\infty}^0  h_p(F_X(x)) \,\dd x
      + \int_0^\infty (1- h_p(F_X(x)))\,\dd x ,
      \qquad p\in[0,1),
  \end{split}
 \end{align}
 where we recall that \ $\overline F_X(x) = 1 - F_X(x) = \PP(X > x)$, \ $x\in\RR$.
\ Formula \eqref{formula_ESn} is also a special case of Theorem 1 in Fuchs et al.\ \cite{FucSchSch}.
We check that both integrals on the right hand side of \eqref{formula_ESn} are finite.
Namely, we have
 \begin{align*}
  \int_0^\infty (1-h_p(F_X(x)))\,\dd x
   & = \int_0^\infty \left(1-  \left(\frac{F_X(x)-p}{1-p}\right)^n\bone_{[p,1]}(F_X(x))\right)\,\dd x \\
   & = \int_{\{x\geq 0: F_X(x)<p\}} 1\,\dd x
       + \int_{\{x\geq 0: F_X(x)\geq p\}} \left(1-  \left(\frac{F_X(x)-p}{1-p}\right)^n \right)\,\dd x.
 \end{align*}
Note that for each \ $x\in\RR$ \ and \ $p\in[0,1)$, \ we have that \ $F_X(x)\geq p$ \ holds if and only if \ $\VaR_X(p)\leq x$,
 \ where \ $\VaR_X(0):=-\infty$ \ (see, e.g., Dhaene et al.\ \cite[formula (1)]{DhaKukLinTan}).
Hence we get
 \begin{align*}
   &\int_0^\infty (1-h_p(F_X(x)))\,\dd x
    = \int_{\{x\geq 0: \VaR_X(p) > x\}} 1\,\dd x
         + \int_{\{x\geq 0: \VaR_X(p)\leq x\}} \left(1-  \left(\frac{F_X(x)-p}{1-p}\right)^n \right)\,\dd x \\
   &\qquad\leq \max(0,\VaR_X(p)) + \int_{\max(0,\VaR_X(p))}^\infty \left(1-  \left(\frac{F_X(x)-p}{1-p}\right)^n \right)\,\dd x \\
   &\qquad = \max(0,\VaR_X(p)) + \int_{\max(0,\VaR_X(p))}^\infty \left(1-  \left(\frac{F_X(x)-p}{1-p}\right) \right)
                                                          \sum_{k=0}^{n-1} \left(\frac{F_X(x)-p}{1-p}\right)^k\,\dd x\\
   &\qquad \leq  \max(0,\VaR_X(p)) + n \int_{\max(0,\VaR_X(p))}^\infty  \left(1-  \left(\frac{F_X(x)-p}{1-p}\right) \right) \,\dd x\\
   &\qquad = \max(0,\VaR_X(p)) + \frac{n}{1-p} \int_{\max(0,\VaR_X(p))}^\infty \left(1-  F_X(x) \right) \,\dd x.
 \end{align*}
Using that \ $\PP(X>x) = \PP(X^+>x)$, \ $x\geq 0$, \ we have
 \begin{align*}
   \int_0^\infty (1-h_p(F_X(x)))\,\dd x
             &\leq \max(0,\VaR_X(p)) + \frac{n}{1-p} \int_{\max(0,\VaR_X(p))}^\infty \PP(X^+>x) \,\dd x\\
             &\leq \max(0,\VaR_X(p)) + \frac{n}{1-p} \int_0^\infty \PP(X^+>x) \,\dd x \\
             &= \max(0,\VaR_X(p)) + \frac{n}{1-p} \EE(X^+)<\infty.
 \end{align*}
Further, using that \ $h_p$ \ is monotone increasing, we get
 \begin{align*}
  \int_{-\infty}^0  h_p(F_X(x)) \,\dd x
    &= \int_{-\infty}^0  h_p(F_X(x)) \bone_{[p,1]}(F_X(x))\,\dd x
     = \int_{-\infty}^0  h_p(F_X(x)) \bone_{(-\infty,0)\cap[\VaR_X(p),\infty)}(x)\,\dd x\\
    & \leq h_p(F_X(0)) \int_{-\infty}^0 \bone_{(-\infty,0)\cap[\VaR_X(p),\infty)}(x)\,\dd x\\
    &  \leq h_p(F_X(0)) \max(0,-\VaR_X(p))
        <\infty.
  \end{align*}
Consequently, we have \ $\ES_{X,n}(p)<\infty$, \ as desired.
Note that the above argument also works under the condition \ $\EE(X^+)<\infty$ \ instead of \ $X\in L^1$.
\proofend

\begin{Rem}
Lemma \ref{Lem_ESn_vegesseg} is in fact an immediate consequence of Proposition 1 in Wang et al.\ \cite{WanWanWei}.
Indeed, for any \ $p\in[0,1)$, \ the distortion function \ $\overline h_p$ \ defined in the proof of Lemma \ref{Lem_ESn_vegesseg}
 is absolutely continuous (in particular, is of bounded variation),
 \ $\overline h_p(0)=0$ \ and the derivative of \ $\overline h_p$ \ (in Lebesgue a.e.\ sense) is bounded.
Further, taking into account the representation \eqref{formula_ESn} of the $n^{\mathrm th}$-order Expected Shortfall,
 we have that the $n^{\mathrm th}$-order Expected Shortfall can be written as formula (2.1) in Wang et al.\ \cite{WanWanWei}
 with the function $\overline h_p$.
Hence one can apply part (i) of Proposition 1 in Wang et al.\ \cite{WanWanWei}.
Our proof of Lemma \ref{Lem_ESn_vegesseg} is different from that of part (i) of Proposition 1 in Wang et al.\ \cite{WanWanWei}.
\proofend
\end{Rem}

\begin{Lem}\label{Lem_pelvehofolytonos_monoton}
Let \ $X$ \ be a random variable such that \ $X\in L^1$, \ and let \ $n\in\NN$.
\ Then the function \ $[0,1)\ni p\mapsto \ES_{X,n}(p)$ \ is continuous and monotone increasing.
\end{Lem}

\noindent{\bf Proof.}
First, we prove the continuity.
Let \ $p\in[0,1)$ \ and \ $(p_m)_{m\in\NN}$ \ be a sequence in \ $[0,1)$ \ such that \ $\lim_{m\to\infty} p_m = p$.
\ We need to check that \ $\lim_{m\to\infty} \ES_{X,n}(p_m) = \ES_{X,n}(p)$.
\ This follows by dominated convergence theorem, since
 \begin{enumerate}
   \item[$\bullet$] if \ $n=1$, \ then \ $\frac{1}{1-p_m}\bone_{[p_m,1]}(s)\to \frac{1}{1-p}\bone_{[p,1]}(s)$ \ as \ $m\to\infty$ \
                    for \ $s\in[0,1]$ \ possibly except \ $s=p$; \ and if \ $n\geq 2$, \ $n\in\NN$, \ then
                    for each \ $s\in[0,1]$, \ we have
                     \[
                     \frac{n}{1-p_m} \left(\frac{s-p_m}{1-p_m}\right)^{n-1} \bone_{[p_m,1]}(s)
                       \to \frac{n}{1-p} \left(\frac{s-p}{1-p}\right)^{n-1} \bone_{[p,1]}(s)
                       \qquad \text{as \ $m\to\infty$,}
                     \]
   \item[$\bullet$] for \ $m\in\NN$ \ satisfying \ $\frac{p}{2}<p_m< p + \frac{1-p}{2} = \frac{1+p}{2}$, \ we have
                    \[
                       \frac{n}{1-p_m}\left(\frac{s-p_m}{1-p_m}\right)^{n-1} \bone_{[p_m,1]}(s)
                       \leq \frac{2n}{1-p}\bone_{\left[\frac{p}{2},1\right]}(s),
                       \qquad s\in [0,1],
                    \]
                    where we used that \ $[p_m,1] \subset \left[\frac{p}{2},1\right]$ \ and that
                    \begin{align*}
                    p_m<\frac{1+p}{2}\;\; \Leftrightarrow \;\; \frac{1}{1-p_m}< \frac{2}{1-p},
                    \end{align*}
   \item[$\bullet$] using that \ $X\in L^1$, \ we have
             \[
              \int_0^1 \bone_{\left[\frac{p}{2},1\right] }(s) \VaR_X(s) \,\dd s
                 = \int_{\frac{p}{2}}^1 \VaR_X(s) \,\dd s
                 = \left(1-\frac{p}{2} \right)\ES_X\left(\frac{p}{2}\right) < \infty.
             \]
 \end{enumerate}

The monotone increasing property of \ $[0,1)\ni p\mapsto \ES_{X,n}(p)$ \ is a consequence of
 Fuchs et al.\ \cite[Corollary 4, part (1)]{FucSchSch}, since if \ $0\leq p_1<p_2<1$, \ then
 \[
   \left(\frac{s-p_2}{1-p_2}\right)^n \bone_{[p_2,1]}(s) \leq  \left(\frac{s-p_1}{1-p_1}\right)^n \bone_{[p_1,1]}(s),\qquad s\in[0,1].
 \]
The monotone increasing property in question also follows by part (i) of Proposition 2 in Wang et al.\ \cite{WanWanWei} and by \eqref{formula_ESn},
 since \ $\overline h_p(s)\geq 1-\frac{s-p}{1-p}\bbone_{[p,1]}(s)$, \ $s\in[0,1]$.
\proofend

Next, we provide some sufficient conditions under which \ $[0,1-\vare) \ni p \mapsto \ES_{X,n}(p)$ \ is strictly monotone increasing,
 where \ $\vare\in(0,1)$ \ and \ $n\in\NN$.
\ Such a result is known in case of \ $n=1$, \ see, e.g., the proof of Proposition 2 in Li and Wang \cite{LiWan}.

\begin{Lem}\label{Lem_szigmonHO}
Let \ $X$ \ be a random variable such that \ $X\in L^1$, \ let \ $\vare\in(0,1)$, \ $n\in\NN$, \ and let us suppose that
  the function \ $(0,1)\ni p \mapsto \VaR_X(p)$ \ is not constant on the interval \ $[1-\vare,1)$.
\ Then the function \ $[0,1-\vare] \ni p \mapsto \ES_{X,n}(p)$ \ is strictly monotone increasing.
\end{Lem}

\noindent{\bf Proof.}
For each \ $p\in[0,1)$, \ let us consider the function \ $h_p$ \ defined in \eqref{Def_hp}.
For each \ $0\leq p_1<p_2<1$, \ we have
 \begin{align}\label{help_hp}
   h_{p_2}(s)\leq h_{p_1}(s),\qquad s\in[0,1];
   \qquad \text{and} \qquad
   h_{p_2}(s)< h_{p_1}(s),\qquad s\in[p_2,1).
 \end{align}
Recall that for each \ $x\in\RR$ \ and \ $p\in[0,1)$, \ we have that \ $F_X(x)\geq p$ \ holds if and only if \ $\VaR_X(p)\leq x$,
 \ where \ $\VaR_X(0):=-\infty$ \ (see, e.g., Dhaene et al.\ \cite[formula (1)]{DhaKukLinTan}).

First, let us consider the case \ $\VaR_X(p_2)\geq 0$.
\ Using \eqref{formula_ESn}, for each \ $0\leq p_1<p_2\leq 1-\vare$, \ we have
 \begin{align*}
   \ES_{X,n}(p_1)
      &= \int_{\{x\geq 0: x<\VaR_X(p_2)\}} (1- h_{p_1}(F_X(x)))\,\dd x
         + \int_{\{x\geq 0: x\geq \VaR_X(p_2)\}} (1- h_{p_1}(F_X(x)))\,\dd x \\
      &\phantom{=\;}   -\int_{-\infty}^0  h_{p_1}(F_X(x)) \,\dd x ,
 \end{align*}
 where, by \eqref{help_hp},
 \[
  \int_{\{x\geq 0: x<\VaR_X(p_2)\}} (1- h_{p_1}(F_X(x)))\,\dd x
    \leq \int_{\{x\geq 0: x<\VaR_X(p_2)\}} (1- h_{p_2}(F_X(x)))\,\dd x,
 \]
 and
 \[
   -\int_{-\infty}^0  h_{p_1}(F_X(x)) \,\dd x
     \leq
    -\int_{-\infty}^0  h_{p_2}(F_X(x)) \,\dd x .
 \]
Further, we have
 \begin{align*}
   \int_{\{x\geq 0: x\geq \VaR_X(p_2)\}} (1- h_{p_1}(F_X(x)))\,\dd x
    & = \int_{\{x\geq 0: p_2\leq F_X(x)<1\}} (1- h_{p_1}(F_X(x)))\,\dd x\\
    & < \int_{\{x\geq 0: p_2\leq F_X(x)<1\}} (1- h_{p_2}(F_X(x)))\,\dd x,
 \end{align*}
 where we used \eqref{help_hp}, the fact that \ $h_p(1)=1$, \ $p\in[0,1)$, \ and that the assumption that the function
  \ $(0,1)\ni p \mapsto \VaR_X(p)$ \ is not constant on the interval \ $[1-\vare,1)$ \
  yields that the set \ $\{x\geq 0: p_2\leq F_X(x)<1\}$ \ has a positive Lebesgue measure.
Consequently, we get
 \begin{align*}
     \ES_{X,n}(p_1)
      &<  \int_{\{x\geq 0: x<\VaR_X(p_2)\}} (1- h_{p_2}(F_X(x)))\,\dd x
         + \int_{\{x\geq 0: x\geq \VaR_X(p_2)\}} (1- h_{p_2}(F_X(x)))\,\dd x \\
      &\phantom{=\;}   -\int_{-\infty}^0  h_{p_2}(F_X(x)) \,\dd x
       = \ES_{X,n}(p_2),
 \end{align*}
 as desired.

Next, let us consider the case \ $\VaR_X(p_2)<0$.
\ Then, similarly as before, using \eqref{help_hp}, we have
 \begin{align*}
   \ES_{X,n}(p_1)
      &\leq \int_0^\infty (1- h_{p_2}(F_X(x)))\,\dd x
         - \int_{\{x\leq 0: F_X(x)=1\}} h_{p_1}(F_X(x)) \,\dd x \\
      &\phantom{=\;}   -\int_{\{x\leq 0: x<\VaR_X(p_2), F_X(x)<1\}}  h_{p_1}(F_X(x)) \,\dd x \\
      &\phantom{=\;}   -\int_{\{x\leq 0: x\geq \VaR_X(p_2), F_X(x)<1\}}  h_{p_1}(F_X(x)) \,\dd x .
 \end{align*}
Here, using again \eqref{help_hp} and that \ $h_p(1)=1$, \ $p\in[0,1)$, \ we have
 \[
   \int_{\{x\leq 0: F_X(x)=1\}} h_{p_1}(F_X(x)) \,\dd x
    = \int_{\{x\leq 0: F_X(x)=1\}} 1 \,\dd x
    = \int_{\{x\leq 0: F_X(x)=1\}} h_{p_2}(F_X(x)) \,\dd x,
 \]
 and
 \[
   -\int_{\{x\leq 0: x<\VaR_X(p_2), F_X(x)<1\}}  h_{p_1}(F_X(x)) \,\dd x
   \leq -\int_{\{x\leq 0: x<\VaR_X(p_2), F_X(x)<1\}}  h_{p_2}(F_X(x)) \,\dd x.
 \]
Further, since the assumption that the function
  \ $(0,1)\ni p \mapsto \VaR_X(p)$ \ is not constant on the interval \ $[1-\vare,1)$ \
  yields that the set
  \[
    \{x\leq 0: x\geq \VaR_X(p_2), F_X(x)<1\}
     = \{x\leq 0: p_2\leq F_X(x)<1\}
  \]
 has a positive Lebesgue measure, by \eqref{help_hp}, we have
 \[
  -\int_{\{x\leq 0: x\geq \VaR_X(p_2), F_X(x)<1\}}  h_{p_1}(F_X(x)) \,\dd x
   <
  -\int_{\{x\leq 0: x\geq \VaR_X(p_2), F_X(x)<1\}}  h_{p_2}(F_X(x)) \,\dd x.
 \]
Consequently, we have \ $\ES_{X,n}(p_1)<\ES_{X,n}(p_2)$, \ as desired.
\proofend

\begin{Def}\label{Def_com_mon}
The random variables \ $X$ \ and \ $Y$ \ are called comonotonic if there exist a random variable \ $Z$ \ and monotone increasing
functions \ $f,g:\RR\to\RR$ \ such that \ $X=f(Z)$ \ and \ $Y=g(Z)$.
\end{Def}

\begin{Pro}\label{Pro_ESn_com_add}
Let \ $X$ \ and \ $Y$ \  be comonotonic random variables such that \ $X,Y\in L^1$, \ and let \ $n\in\NN$.
\ Then
 \[
  \ES_{X+Y,n}(p) = \ES_{X,n}(p) + \ES_{Y,n}(p), \qquad p\in[0,1),
 \]
 that is, the $n^{\mathrm {th}}$-order Expected Shortfall is additive for comonotonic random variables belonging to $L^1$.
\end{Pro}

Proposition \ref{Pro_ESn_com_add} is the direct consequence of the additivity of \ $\VaR$ \ for comonotonic random variables
 (see, e.g., McNeil et al.\ \cite[Proposition 7.20]{McnFreEmb}) and the additivity of Lebesgue integral.

Next, we investigate the connection between weak convergence of random variables and the convergence of their
 higher-order Expected Shortfalls.

\begin{Lem}\label{Lem_ES2_conv}
Let \ $n\in\NN$, \ $X_m$, $m\in\NN$, \ and \ $X$ \ be random variables such that \ $X_m\in L^1$, \ $m\in\NN$, \ and \ $X\in L^1$.
\ If \ $X_m\distr X$ \ as \ $m\to\infty$, \ and \ $\{X_m : m\in\NN\}$ \ is uniformly integrable, then \ $\ES_{X_m,n}(p)\to \ES_{X,n}(p)$ \ as \ $m\to\infty$
 \ for each \ $p\in[0,1)$.
\end{Lem}

\noindent{\bf Proof.}
For each random variable \ $Y\in L^1$, \ we have
 \begin{align}\label{help_ES_13}
 \begin{split}
  \ES_{Y,n}(p) & = \frac{n}{(1-p)^n} \int_0^1 (s-p)^{n-1}\VaR_Y(s)\bone_{[p,1]}(s)\,\dd s \\
               & = \frac{n}{(1-p)^n} \EE\big( (U-p)^{n-1} \VaR_Y(U) \bone_{[p,1]}(U) \big),
               \qquad p\in[0,1),
 \end{split}
 \end{align}
 where \ $U$ \ is a uniformly distributed random variable on \ $(0,1)$.
\ Recall also that if \ $Y_m$, \ $m\in\NN$, \ and \ $Y$ \ are random variables such that \ $Y_m\distr Y$ \ as \ $m\to\infty$ \ and
 \ $\{Y_m : m\in\NN\}$ \ is uniformly integrable, then \ $\EE(\vert Y\vert)<\infty$ \ (i.e., \ $Y\in L^1$) \ and \ $\EE(Y_m)\to \EE(Y)$ \ as \ $m\to\infty$,
 \ see, e.g., Billingsley \cite[Theorem 5.4]{Bil}.

By \eqref{help_ES_13}, for each \ $m\in\NN$, \ we have
 \[
  \ES_{X_m,n}(p)  = \frac{n}{(1-p)^n} \EE\big( (U-p)^{n-1} \VaR_{X_m}(U) \bone_{[p,1]}(U) \big), \qquad p\in[0,1),
 \]
 and
 \[
  \ES_{X,n}(p)  = \frac{n}{(1-p)^n} \EE\big( (U-p)^{n-1} \VaR_X(U) \bone_{[p,1]}(U) \big), \qquad p\in[0,1).
 \]
Hence to prove the statement, it is enough to verify that for each \ $p\in[0,1)$,
 \begin{itemize}
   \item[(i)] $(U-p)^{n-1} \VaR_{X_m}(U) \bone_{[p,1]}(U) \distr (U-p)^{n-1} \VaR_X(U) \bone_{[p,1]}(U)$ \ as \ $m\to\infty$,
   \item[(ii)] the family \ $\Big\{ (U-p)^{n-1} \VaR_{X_m}(U) \bone_{[p,1]}(U) : m\in\NN\Big\}$ \ is uniformly integrable,
 \end{itemize}
 where \ $U$ \ is a uniformly distributed random variable on \ $(0,1)$.

In what follows let \ $p\in[0,1)$ \ be fixed.
Since \ $X_m\distr X$ \ as \ $m\to\infty$, \ by the quantile convergence
 theorem (see, e.g., Shorack and Wellner \cite[Exercise 5, page 10]{ShoWel}), we have
 \ $\VaR_{X_m}(q)\to \VaR_X(q)$ \ as \ $m\to\infty$ \ for each continuity point \ $q\in(0,1)$ \ of the function \ $(0,1)\ni z\mapsto \VaR_X(z)$.
\ Since function \ $(0,1)\ni z\mapsto \VaR_X(z)$ \ is monotone increasing, it has at most countable many discontinuity points.
Hence using that \ $U$ \ is absolutely continuous, we have \ $\VaR_{X_m}(U)$ \ converges to \ $\VaR_X(U)$ \ as \ $m\to\infty$ \ almost surely, yielding (i).
Further, we have
 \[
   \Big\vert (U-p)^{n-1} \VaR_{X_m}(U) \bone_{[p,1]}(U) \Big\vert
      \leq \vert \VaR_{X_m}(U)\vert,\qquad m\in\NN,
 \]
 and \ $\VaR_{X_m}(U)$ \ has the same distribution as \ $X_m$ \ for each \ $m\in\NN$ \ (see, e.g., Embrechts and Hofert
 \cite[Proposition 2]{EmbHof}).
Consequently, the uniform integrability of the family \ $\{X_m : m\in\NN\}$ \ yields
  that of the family \ $\{\VaR_{X_m}(U) : m\in\NN\}$ \ (see, e.g., Billingsley \cite[page 32]{Bil}).
Hence we get (ii), as desired.
\proofend

\begin{Rem}
Lemma \ref{Lem_ES2_conv} does in fact follow from a more general result of Wang et al.\ \cite[Theorem 6]{WanWanWei}.
Indeed, for each \ $p\in[0,1)$, \ one can use \eqref{formula_ESn}, where the distortion function \ $\overline h_p$ \
 is continuous, of bounded variation with \ $\overline h_p(0)=1$, \ and we check that the family \ $\{X,X_m, m\in\NN\}$ \
 is \ $\overline h_p$-uniformly integrable in the sense
 of Wang et al.\ \cite[Section 4]{WanWanWei} provided that \ $X\in L^1$ \ and \ $\{X_m, m\in\NN\}$ \ is uniformly integrable.
Let us suppose that \ $X\in L^1$ \ and \ $\{X_m,m\in\NN\}$ \ is uniformly integrable.
Then we readily have that \ $\{X,X_m,m\in\NN\}$ \ is uniformly integrable as well.
Further, for each \ $p\in[0,1)$ \ and \ $k\in(0,1-p)$, \ we have
 \begin{align}\label{help_uniform}
  \begin{split}
   \int_0^k \vert \VaR_Y(1-s)\vert\,\dd \overline h_p(s)
    & = \int_0^k \vert \VaR_Y(1-s)\vert h_p'(1-s)\,\dd s\\
    & = \int_0^k \vert \VaR_Y(1-s)\vert \frac{n}{(1-p)^n}(1-s-p)^{n-1} \,\dd s\\
    & = \frac{n}{(1-p)^n} \EE\Big( \vert \VaR_Y(1-U)\vert (1-U -p)^{n-1} \bone_{\{U<k\}}\Big)\\
    &  = \frac{n}{(1-p)^n} \EE\Big( \vert \VaR_Y(U)\vert (U -p)^{n-1} \bone_{\{U>1-k\}}\Big) \\
    & \leq \frac{n}{(1-p)^n} \EE\Big( \vert \VaR_Y(U)\vert \bone_{\{U>1-k\}}\Big)
  \end{split}
 \end{align}
 for each \ $Y\in\{X,X_m,m\in\NN\}$, \ where \ $U$ \ is a uniformly distributed random variable on
 \ $(0,1)$, \ and hence \ $1-U$ \ is uniformly distributed on \ $(0,1)$ \ as well.
Recall also that \ $\VaR_Y(U)$ \ has the same distribution as \ $Y$ \ for \ $Y\in\{X,X_m,m\in\NN\}$ \
 (see, e.g., Embrechts and Hofer \cite[Proposition 2]{EmbHof}).
Consequently, the family \ $\{\VaR_X(U), \VaR_{X_m}(U), m\in\NN\}$ \ is uniformly integrable as well.
Hence for each \ $\vare>0$, \ there exists \ $\delta>0$ \ such that \ $\EE(\vert \VaR_Y(U)\vert\bone_A)<\vare$ \
 for each \ $Y\in\{X,X_m,m\in\NN\}$ \ and for each event \ $A\in \cF$ \ with \ $\PP(A)<\delta$.
\ By choosing \ $A:=\{U>1-k\}$, \ for each \ $\vare>0$ \ there exists \ $\delta>0$ \ such that for each \ $k\in(0,\delta)$, \
 we have
 \[
   \sup_{Y\in\{X,X_m,m\in\NN\}} \EE(\vert \VaR_Y(U)\vert\bone_{\{U>1-k\}})<\vare.
 \]
Taking into account \eqref{help_uniform}, it yields that
 \begin{align}\label{help_uni_int1}
  \lim_{k\downarrow 0} \sup_{Y\in\{X,X_m,m\in\NN\}} \int_0^k \vert \VaR_Y(1-s)\vert\,\dd \overline h_p(s) = 0.
 \end{align}

Further, for each \ $p\in(0,1)$ \ and \ $k\in(1-p,1)$, \ we have
 \begin{align*}
   \int_k^1 \vert \VaR_Y(1-s)\vert\,\dd \overline h_p(s)
    = \int_k^1 \vert \VaR_Y(1-s)\vert h_p'(1-s)\,\dd s
    = \int_k^1 \vert \VaR_Y(1-s)\vert \cdot 0 \,\dd s
    = 0
 \end{align*}
 for \ $Y\in\{X,X_m,m\in\NN\}$, \ yielding that
 \begin{align}\label{help_uni_int2}
   \lim_{k\uparrow 1} \sup_{Y\in\{X,X_m,m\in\NN\}} \int_k^1 \vert \VaR_Y(1-s)\vert\,\dd \overline h_p(s) = 0.
 \end{align}
Similarly, as we have seen at the beginning of the remark, one can check that
 \begin{align}\label{help_uni_int3}
   \lim_{k\uparrow 1} \sup_{Y\in\{X,X_m,m\in\NN\}} \int_k^1 \vert \VaR_Y(1-s)\vert\,\dd \overline h_0(s) = 0
 \end{align}
 holds as well (instead of \ $A=\{U>1-k\}$ \ one can choose \ $A=\{U<1-k\}$).
By \eqref{help_uni_int1}, \eqref{help_uni_int2} and \eqref{help_uni_int3}, we get that
 the family \ $\{X,X_m, m\in\NN\}$ \ is \ $\overline h_p$-uniformly integrable in the sense
 of Wang et al.\ \cite[Section 4]{WanWanWei}, as desired.

Finally, note that the intrinsic reason for the fact that the \ $\overline h_p$-uniformly integrable of \ $\{X,X_m, m\in\NN\}$ \
 follows from the uniform integrability of \ $\{X,X_m, m\in\NN\}$ \ is that \ $\overline h_p'$ \ (in Lebesgue a.e. sense) is non-negative and bounded.
\proofend
\end{Rem}

\section{Second-order Expected Shortfall and Gini Shortfall}\label{App_2ndES_GiniES}

Let \ $X$ \ be a random variable such that \ $X\in L^1$, \ and let \ $p\in[0,1)$.
\ Let \ $U_p$ \ be a random variable uniformly distributed on the interval \ $(p,1)$, \
 and let \ $F_{X,p}$ \ be the distribution function of the random variable \ $\VaR_{X}(U_p)$.
\ The tail-Gini functional of \ $X$ \ at a level \ $p\in[0,1)$ \ is defined by
 \[
     \TGINI_X(p):= \EE(\vert X_p^* - X_p^{**}\vert),
 \]
 where \ $X_p^*$ \ and \ $X_p^{**}$ \ are two independent, identically distributed random variables
 with a distribution function \ $F_{X,p}$, \ see Furman et al.\ \cite[formula (3.6)]{FurWanZit}.
\ Note that if \ $p=0$, \ then \ $\VaR_X(U_0)$ \ has the same distribution as \ $X$ \ (see, e.g., Embrechts and Hofert \cite[Proposition 2]{EmbHof}),
 and hence \ $\TGINI_X(0)$ \ is nothing else but the Gini variability measure of \ $X$ \ given by
 \ $\EE(\vert X^* -  X^{**} \vert)$, \ where \ $X^*$ \ and \ $X^{**}$ \ are two independent copies of \ $X$.
\ By Furman et al.\ \cite[(3.3) and Proposition 3.2]{FurWanZit}, we have
 \[
   \TGINI_X(p) = \frac{2}{(1-p)^2}\int_p^1 (2s - 1 - p)\VaR_X(s)\,\dd s,\qquad p\in[0,1).
 \]

For \ $\lambda\geq 0$, \ the Gini Shortfall of \ $X$ \ at a level \ $p\in[0,1)$ \ corresponding to the (loading)
 parameter \ $\lambda$ \ is defined by
 \[
     \GS_X(p,\lambda):= \ES_X(p) + \lambda \TGINI_X(p),
 \]
 see Furman et al.\ \cite[formula (4.1)]{FurWanZit}.
By Theorem 4.1 in Furman et al.\ \cite{FurWanZit},
 \begin{align}\label{help_GS_1}
   \GS_X(p, \lambda) = \frac{1}{(1-p)^2}\int_p^1 \left(1-p + 4\lambda\left(s-\frac{1+p}{2}\right)\right)\VaR_X(s)\,\dd s,
                      \qquad p\in[0,1),\;\; \lambda\geq 0,
 \end{align}
 and, using also Lemma 4.2 in Furman et al.\ \cite{FurWanZit}, the Gini Shortfall at a level \ $p$ \ corresponding to the parameter \ $\lambda$ \
 is a coherent risk measure on \ $L^1$ \ if and only if \ $\lambda\in[0,\frac{1}{2}]$.

By \eqref{help_GS_1}, we have
 \begin{align*}
  \GS_X\left(p,\frac{1}{2}\right)
                & = \frac{1}{(1-p)^2}\int_p^1 \left(1-p + 2\left(s-\frac{1+p}{2}\right)\right)\VaR_X(s)\,\dd s\\
                & =  \frac{2}{(1-p)^2}\int_p^1 (s-p) \VaR_X(s)\,\dd s
                 = \ES_{X,2}(p), \qquad p\in[0,1),
 \end{align*}
 so for \ $p\in[0,1)$, \ the \ $2^{\mathrm{nd}}$-order Expected Shortfall of \ $X$ \ at a level \ $p$ \ is nothing else but
 the Gini Shortfall of \ $X$ \ at a level \ $p$ \ corresponding to the parameter \ $\frac{1}{2}$.
In particular, we have that the \ $2^{\mathrm{nd}}$-order Expected Shortfall (at any level \ $p\in[0,1)$) \
 is a coherent risk measure on \ $L^1$ \ (which is in accordance with part (ii) of Remark \ref{Rem4}).

Further, using again \eqref{help_GS_1}, for each \ $p\in[0,1)$ \ and \ $\lambda\geq 0$, \ we have
 \begin{align*}
  \GS_X(p,\lambda)
    &  = \frac{1}{(1-p)^2} \int_p^1 \left(1-p + 4\lambda\left(s-p + \frac{p-1}{2}\right) \right)\VaR_X(s)\,\dd s\\
    & = (1-2\lambda) \frac{1}{1-p}\int_p^1 \VaR_X(s)\,\dd s
       + 2\lambda \frac{2}{(1-p)^2}\int_p^1 (s-p)\VaR_X(s)\,\dd s\\
    & = (1-2\lambda)\ES_X(p) + 2\lambda \ES_{X,2}(p),
 \end{align*}
 yielding that the Gini Shortfall at a level \ $p\in[0,1)$ \ corresponding to a parameter \ $\lambda\geq 0$ \ is
 the linear combination of the Expected Shortfall at level \ $p$ \ and the \ $2^{\mathrm{nd}}$-order Expected Shortfall at level \ $p$
 \ with coefficients \ $1-2\lambda$ \ and \ $2\lambda$, \ respectively.

\section{Karamata theorem for regularly varying functions at zero}\label{Sec_Karamata_zero}

For the notions of a regularly varying function at infinity and at $0$, respectively, see Definition \ref{Def_reg_var}.

We formulate a Karamata theorem for regularly varying functions at \ $0$ \ with index \ $\kappa>-1$,
 \ which is used in the proof of Theorem \ref{Thm_regularly_varying}.
We could not address any reference for it, and hence, for completeness, we provide a proof as well.

\begin{Lem}\label{Lem_Karamata_zero}
Let \ $x_0>0$ \ and \ $f:(0,x_0)\to(0,\infty)$ \ be a regularly varying function at \ $0$ \ with index \ $\kappa>-1$.
\ Then
 \[
    \lim_{\vare\downarrow 0} \frac{\int_0^\vare f(v)\,\dd v}{\vare f(\vare)}
       = \frac{1}{\kappa +1}.
 \]
\end{Lem}

\noindent{\bf Proof.}
For each \ $\vare\in(0,x_0)$, \ by the substitution \ $v=\frac{1}{u}$, \ we have
 \begin{align}\label{help_Karamata_zero}
  \int_0^\vare f(v)\,\dd v = \int_{\frac{1}{\vare}}^\infty u^{-2} f\left(\frac{1}{u}\right)\,\dd u.
 \end{align}
Here the function \ $(\frac{1}{x_0},\infty)\ni u\mapsto u^{-2} f\left(\frac{1}{u}\right)$ \ is regularly varying at infinity
 with index \ $-\kappa-2<-1$ \ (see Definition \ref{Def_reg_var}), since it is measurable and
 \[
 \lim_{u\to\infty} \frac{(qu)^{-2} f\left(\frac{1}{qu}\right)}{u^{-2} f\left(\frac{1}{u}\right)}
     = q^{-2} \lim_{\vare\downarrow 0} \frac{f\left(\frac{1}{q}\vare\right)}{f(\vare)}
     = q^{-2} \left(\frac{1}{q}\right)^{\kappa}
     = q^{-\kappa-2}\qquad \text{for each \ $q>0$.}
 \]
Consequently, by Karamata theorem for regularly varying functions at infinity with index strictly less than \ $-1$ \
 (see, e.g., Resnick \cite[Theorem 0.6]{Res}), we get
 \[
   \lim_{\vare\downarrow 0} \frac{\int_{\frac{1}{\vare}}^\infty u^{-2} f\left(\frac{1}{u}\right)\,\dd u}{\frac{1}{\vare}\cdot \left(\frac{1}{\vare}\right)^{-2}f(\vare)}
      = \frac{1}{\kappa + 2 -1}
      = \frac{1}{\kappa + 1}.
 \]
Hence, by \eqref{help_Karamata_zero}, the assertion follows.
\proofend

\section*{Acknowledgements}
We would like to thank the referees for their comments that helped us improve the paper.

\end{document}